\documentclass{article}

\usepackage[a4paper,total={6.1in, 8.2in}]{geometry}
\usepackage{algorithm}
\usepackage{algorithmic}

\usepackage{bm,amsmath,amssymb,amsfonts,graphicx,epsfig,amsthm,color}
\usepackage{dsfont,bbold}
\usepackage{caption}

\usepackage[hyphens]{url}
\usepackage{hyperref}
\usepackage{url}            % simple URL typesetting
\usepackage{booktabs}       % professional-quality tables
\usepackage{amsfonts}       % blackboard math symbols
\usepackage{nicefrac}       % compact symbols for 1/2, etc.
\usepackage{microtype}      % microtypography

\usepackage{times}
\usepackage{graphicx} % more modern
\usepackage{subfigure} 

\usepackage{wrapfig}

\newcommand{\RN}[1]{%
  \textup{\uppercase\expandafter{\romannumeral#1}}%
}

\usepackage{accents}
\makeatletter
\def\wid{\check{{\cc@style\underline{\mskip9.5mu}}}}
\def\Wideubar{\underaccent{{\cc@style\underline{\mskip6mu}}}}
\makeatother

\makeatletter
\def\wideubar{\underaccent{{\cc@style\underline{\mskip9.5mu}}}}
\def\Wideubar{\underaccent{{\cc@style\underline{\mskip6mu}}}}
\makeatother

\makeatletter
\def\widebar{\accentset{{\cc@style\underline{\mskip9.5mu}}}}
\def\Widebar{\accentset{{\cc@style\underline{\mskip6mu}}}}
\makeatother

\newtheorem{proposition}{Proposition}

\newtheorem{theorem}{Theorem}

\theoremstyle{remark}

\newcommand{\minimize}{{\rm minimize}}

\def\ccalH{{\ensuremath{\mathcal H}}}
\def\ccalT{{\ensuremath{\mathcal T}}}

\begin{document}
%\bstctlcite{IEEEexample:BSTcontrol}
\title{%Scalable Solvers of Random Quadratic Equations via Stochastic Truncated Amplitude Flow
Solving Large-scale Systems of Random Quadratic Equations via Stochastic Truncated Amplitude Flow
%with $\mathcal{O}(n)$ Complexities
}

%\author{Gang Wang,~\emph{Student Member,~IEEE}, Bo Yang,~\emph{Student Member,~IEEE,}\\
% Georgios B. Giannakis,~\emph{Fellow,~IEEE,} and Nikolaos D. Sidiropoulos,~\emph{Fellow,~IEEE}
%\thanks{G. Wang, B. Yang, G. B. Giannakis, and N. D. Sidiropoulos are with Department of Electrical and Computer Engineering, University of Minnesota, Minneapolis, MN 55455. G. Wang is also  with School of Automation, Beijing Institute of Technology, Beijing, China, 100081. Emails: \{gangwang,yang4173,georgios,nikos\}@umn.edu.} 
%%  \\
%%  1. Department of Electrical and Computer Engineering\\
%%  University of Minnesota\\
%%  Minneapolis, MN 55455 \\
%%  2. School of Automation\\
%%  Beijing Institute of Technology\\
%%  Beijing, China, 100081\\
%%    \texttt{\{gangwang,\,georgios\}@umn.edu} \\
%%3. EE Dept., Technion -\\
%%Israel Institute of Technology\\
%%Haifa 32000, Israel\\
%%\texttt{yonina@ee.technion.ac.il}
%}

\author{
Gang Wang,~\emph{Student Member,~IEEE}, Georgios B. Giannakis,~\emph{Fellow,~IEEE},\\
and Jie Chen,~\emph{Senior Member,~IEEE}
%	 and
%	Yonina C. Eldar,~\IEEEmembership{Fellow,~IEEE}
%	%\\ May 26, 2016; Revised: July 1, 2016.
%%	\vspace*{-2em} %% REMOVE THIS IN THE REVISED VERSION
%
\thanks{Work in this paper was supported in part by NSF grants 1500713 and 1514056. G. Wang and G. B. Giannakis are with the Digital Technology Center and the ECE Dept., University of Minnesota, Minneapolis, MN 55455, USA. 
G. Wang is also with the School of Automation, Beijing Institute of Technology, Beijing 100081, P. R. China.
J. Chen is with the School of Automation and State Key Laboratory of Intelligent Control and Decision of Complex Systems, Beijing Institute of Technology, Beijing 100081, P. R. China. E-mails: \{gangwang,georgios\}@umn.edu; chenjie@bit.edu.cn.
%G. Wang is with the School of Automation, Beijing Institute of Technology, Beijing 100081, P. R. China, and also with the ECE Dept., University of Minnesota, Minneapolis, MN 55455, USA. G. B. Giannakis is with the Digital Technology Center and the ECE Dept., University of Minnesota, Minneapolis, MN 55455, USA. J. Chen is with the School of Automation, Beijing Institute of Technology, Beijing 100081, P. R. China. E-mails: \{gangwang,georgios\}@umn.edu; chenjie@bit.edu.cn.
}
%
%%\thanks{Color versions of one or more of the figures is this paper are available online at {http://ieeexplore.ieee.org}.}
%
%%\thanks{Digital Object Identifier XXXXXX}
}

%\markboth{}{Wang, Giannakis, and Eldar: Solving  Systems of Random Quadratic Equations via Truncated Amplitude Flow}

\maketitle

\begin{abstract}
A novel approach termed \emph{stochastic truncated amplitude flow} (STAF) is developed to reconstruct an unknown $n$-dimensional real-/complex-valued signal $\bm{x}$ from $m$ `phaseless' quadratic equations of the form $\psi_i=|\langle\bm{a}_i,\bm{x}\rangle|$. This problem, also known as phase retrieval from magnitude-only information, is \emph{NP-hard} in general. Adopting an amplitude-based nonconvex formulation, STAF leads to an iterative solver comprising two stages: s1) Orthogonality-promoting initialization through a stochastic variance reduced gradient algorithm; and, s2) A series of iterative refinements of the initialization using stochastic truncated gradient iterations. Both stages involve a single equation per iteration, thus rendering STAF a simple, scalable, and fast approach amenable to large-scale implementations that is useful when $n$ is large. When $\{\bm{a}_i\}_{i=1}^m$ are independent Gaussian, STAF provably recovers exactly any $\bm{x}\in\mathbb{R}^n$ exponentially fast based on order of $n$ quadratic equations. STAF is also robust in the presence of additive noise of bounded support. Simulated tests involving real Gaussian $\{\bm{a}_i\}$ vectors demonstrate that STAF empirically reconstructs any $\bm{x}\in\mathbb{R}^n$ exactly from about $2.3n$ magnitude-only measurements, outperforming state-of-the-art approaches and narrowing the gap from the information-theoretic number of equations $m=2n-1$. Extensive experiments using synthetic data and real images corroborate markedly improved performance of STAF over existing alternatives.  
\end{abstract}

\begin{keywords}
 Nonconvex optimization, phase retrieval, variance reduction, Kaczmarz algorithm.
\end{keywords}

\section{Introduction}\label{sec:intro}
Consider the fundamental problem of reconstructing a general signal vector from magnitude-only measurements, e.g., the magnitude of the Fourier transform or any linear transform of the signal. This problem, also known as \emph{phase retrieval}~\cite{wf}, arises in many fields of science and engineering ranging from X-ray crystallography~\cite{nature1999miao}, optics~\cite{optics}, as well as coherent diffraction imaging~\cite{diffraction}. %, as well as ptychography~\cite{ptychography}. % to acoustics~\cite{2006balan}, and quantum information~\cite{quantum}. %to astronomy~\cite{astronomy}, and microscopy~\cite{micro}. 
% to acoustics~\cite{2006balan}, blind channel estimation~\cite{blind}, and quantum information~\cite{quantum}. 
In such settings, due to the physical limitations of optical detectors such as photosensitive films, charge-coupled device (CCD) cameras, and human eyes, one records only the intensity of light (which describes the absolute counts of photons or electrons that strike the detectors)  
but loses the phase (where the wave peaks and troughs lie)~\cite{spm2016eldar}. 
%cannot measure the phase of a light wave and record the photon flux instead. 
It is known that when collecting the diffraction pattern at a large enough distance from the imaging plane, the field is given by the Fourier transform of the image (up to a known phase factor).
Therefore, those optical devices in the far field essentially measure only the squared modulus of the Fourier transform of the object,  %Fresnel or Fraunhofer diffraction pattern of the radiation scattered by an object, 
whereas the phase of the incident light reaching the detector is missing.  
Nevertheless, very much information is contained in the Fourier phase. It has been well documented that the Fourier phase of an image encodes often more structural information than its Fourier magnitude~\cite{siam2015candes}. Recovering the phase from magnitude-only measurements is thus of paramount practical relevance. Further details concerning recent advances in the theory and practice of phase retrieval can be found in the review~\cite{spm2016eldar}.  
%Therefore, the problem of uniquely recovering the phase from only the magnitude measurements is ill-posed. 
 
%intensity of the incident light but not the phase. 
Succinctly stated, the generalized phase retrieval amounts to solving a system of `phaseless' quadratic equations taking the form
\begin{equation}\label{eq:quad}
\psi_i = \left|\langle\bm{a}_i,\bm{x}\rangle\right|,\quad 1\le i\le m
\end{equation}
where $\bm{x}\in\mathbb{R}^n$ or $\mathbb{C}^n$ is the wanted unknown, $\bm{a}_i\in\mathbb{R}^n$ or $\mathbb{C}^n$ are known sensing/feature vectors, and $\bm{\psi}:=\left[\psi_1~\cdots~\psi_m\right]^\ccalT$ is the observed data vector. Equivalently, \eqref{eq:quad} can also be given in its squared form as $y_i=|\langle \bm{a}_i,\bm{x}\rangle|^2$, where $y_i:=\psi_i^2$ denotes the intensity or the squared modulus. %In the listed examples, the $\bm{a}_i$'s comprise complex exponentials evaluated at different frequencies $w_i$ so that one collects the (squared) modulus of the Fourier transform of $\bm{x}$. 
%When $\left\{\bm{a}_i\right\}_{i=1}^m$ are complex-valued, the amplitudes of the inner-products $\left\{\langle\bm{a}_i,\bm{x}\rangle\right\}$ are given but the phase information is lacking; yet in the real case only the signs of $\left\{\langle\bm{a}_i,\bm{x}\rangle\right\}$ are unknown. 

In the classical discretized one-dimensional 
%($1$D refers to a vector $\bm{x}$ as opposed to $2$D for a matrix $\bm{X}$) 
phase retrieval, the amplitude vector $\bm{\psi}$ corresponds to the $m$-point (typically, $m= 2n-1$) Fourier transform of the length-$n$ signal $\bm{x}$~\cite{spm2016eldar}.  
It has been established using the fundamental theorem of algebra that there is no unique solution in the discretized $1$D phase retrieval, even if one fixes trivial ambiguities resulting from operations that preserve Fourier magnitudes, including the global phase shift, conjugate inversion, and spatial shift~\cite{1duniqueness}. In fact, there are up to $2^{n-2}$ generally distinct signals with common $\bm{\psi}$ beyond trivial ambiguities % when $(2n-1)$-point Fourier transform is performed~
\cite{1duniqueness}. 
To overcome this ill-posed character of the $1$D phase retrieval, different approaches have been suggested. Additional constraints on the unknown signal such as sparsity or non-negativity are enforced in~\cite{gespar,1982fienup,
altmin,as2016clm,
pr2015qian,sparta}. Other viable options include introducing specific redundancy into measurements leveraging for example, the short-time Fourier transform~\cite{spm2016eldar,arxiv2016be}, or masks
\cite{coded}, or simply assuming
 random measurements (e.g., Gaussian $\{\bm{a}_i\}$ designs)~\cite{altmin,wf,twf,taf}. For analytic concreteness, we will henceforth assume random measurements $\bm{\psi_i}$ that are collected from the real-valued Gaussian model~\eqref{eq:quad}, with independently and identically distributed (i.i.d.) $\bm{a}_i\sim \mathcal{N}(\bm{0},\bm{I}_n)$. To demonstrate the effectiveness of our proposed algorithm, experimental implementation for the complex-valued Gaussian model with i.i.d. $\bm{a}_i\sim \mathcal{CN}(\bm{0},\bm{I}_n):= \mathcal{N}(\bm{0},\bm{I}_n/2)+j\mathcal{N}(\bm{0},\bm{I}_n/2)$, 
 and using real images will be included as well.

It has been recently proved that when $m\ge 2n-1$ or $m\ge 4n-4$ generic measurements (e.g., from the Gaussian models)
are acquired, the system in \eqref{eq:quad} determines uniquely an $n$-dimensional real- or complex-valued $\bm{x}$ (up to a global sign or phase)~\cite{2006balan,4m-4}, respectively.
In the real case, $m=2n-1$ generic measurements are also proved necessary for uniqueness~\cite{2006balan}. 
Postulating existence of a unique solution $\bm{x}$, our goal is to devise simple yet effective algorithms amenable to large-scale implementation: i) that provably reconstruct $\bm{x}$ from a near-optimal number of phaseless quadratic equations as in~\eqref{eq:quad}; and ii), that feature in simultaneously low iteration and computational complexities as well as a linear convergence rate.

Being a particular instance of nonconvex quadratic programming, 
the problem of solving quadratic equations subsumes as special cases various classical combinatorial optimization tasks involving Boolean variables (e.g., the \emph{NP-complete} stone problem~\cite[Section 3.4.1]{siam2001bental},~\cite{twf}).  
%is typically \emph{NP-hard}, and hence computationally intractable~\cite{nphard}. 
%Specifically, consider a real-valued unknown signal $\bm{x}$, and assume that $\bm{x}$ can be uniquely determined by generic measurements $y_i$, i.e., if $\bm{x}$ is a solution to \eqref{eq:quad}, so is $-\bm{x}$. 
Considering for instance the real-valued vectors $\bm{a}_i$ and $\bm{x}$, this problem boils down to assigning signs $s_i=\pm 1$, such that the solution to the system of linear equations $\langle \bm{a}_i,\bm{x}\rangle=s_i\sqrt{y_i}$, denoted by $\bm{z}$, adheres to the given phaseless equations $|\langle \bm{a}_i,\bm{z}\rangle|=\psi_i$, $1\le i\le m$. It is clear that there are a total of $2^m$ different combinations of $\{s_i\}_{i=1}^m$, whereas only two combinations of these signs leads to $\bm{x}$ up to a global sign. The complex scenario becomes even more complicated, in which instead of assigning a series of signs $\{s_i\}_{i=1}^m$, one looks for a collection of unimodular complex constants $\{\sigma_i\in\mathbb{C}\}_{i=1}^m$ such that the resulting linear system and the original quadratic system are equivalent. Furthermore, solving quadratic equations has also found applications in estimating the mixture of linear regressions, in which the latent membership variables are viewed as the missing phases~\cite{mixedlr}. Indeed, it has been shown that reconstructing a discrete-time, finite-duration signal from its Fourier transform magnitudes is \emph{NP-complete}~\cite{nphard1}. Despite its practical relevance across various science and engineering fields, solving systems of quadratic equations is combinatorial in nature, and \emph{NP-hard} in general.
%it is well known that even checking whether a solution exists or not is \emph{NP-complete} as well.  

\emph{Notation}. Lower- (upper-) case boldface letters denote column vectors (matrices), and calligraphic symbols are reserved for sets. The symbol $\ccalT$ ($\ccalH$) stands for transposition (conjugate transposition), and $\succeq$ for positive semidefinite matrices. For vectors, $\|\!\cdot\!\|$ signifies the Euclidean norm; and $\|\!\cdot\!\|_{1}$  denotes the $\ell_1$-norm, respectively. The symbol $\lceil\cdot\rceil$ is the ceiling operation that returns the smallest integer greater than or equal to the given number.
For a given function $g(n)$ of integer $n>0$, $\Theta(g(n))$ denotes the set of functions %~\cite{book2009clrs}
$
		\Theta(g(n))=\big\{f(n)\!:\text{there exist positive constants}~C_1,\,C_2,~\text{and}~n_0~\text{such that}~0\le C_1g(n)\le f(n)\le C_2 g(n)	~
\text{for all} ~n\ge n_0
	\big\}
$; 
and likewise,
$
		\mathcal{O}(g(n))=\big\{f(n)\!:\text{there exist positive constants}~C~\text{and}~n_0$ such that $0\le f(n)\le C g(n)~\text{for all} ~n\ge n_0
	\big\}
$,
and 
$
		\Omega(g(n))=\big\{f(n)\!:\text{there exist positive constants}~C~\text{and}~n_0$ such that $0\le Cg(n)\le f(n)~\text{for all} ~n\ge n_0
	\big\}
$. 

\subsection{Prior Art}

Adopting the least-squares criterion (which would coincide with the maximum likelihood one when assuming an additive white Gaussian noise model), 
the task of tackling the quadratic system in~\eqref{eq:quad} 
can be reformulated as  
that of minimizing the following \emph{amplitude-based} empirical loss~\cite{1982fienup,altmin,taf}
\begin{equation}\label{eq:ls1}
\underset{\bm{z}\in\mathbb{C}^n}{\text{minimize}}~~%\ell(\bm{z}):=
\frac{1}{2m}\sum_{i=1}^m
\left(\psi_i-|\bm{a}_i^\ccalH\bm{z}|\right)^2
\end{equation} 
or, the \emph{intensity-based} one~\cite{wf} %when $\psi_i$ observed instead of $y_i$ in AWGN
\begin{equation}\label{eq:ls}
\underset{\bm{z}\in\mathbb{C}^n}{\text{minimize}}~~%f(\bm{z}):=
\frac{1}{2m}\sum_{i=1}^m\left(y_i-|\bm{a}_i^\ccalH\bm{z}|^2\right)^2
\end{equation}
  and its counterpart for Poisson data~\cite{twf}
\begin{equation}\label{eq:ls2}
\underset{\bm{z}\in\mathbb{C}^n}{\text{minimize}}~~%p(\bm{z}):=
\frac{1}{2m}\sum_{i=1}^m
|\bm{a}_i^\ccalH\bm{z}|^2-y_i\log\big(|\bm{a}_i^\ccalH\bm{z}|^2\big).
\end{equation} 
Unfortunately, the three objective functions are  
 nonconvex because of the modulus in \eqref{eq:ls1}, or the quadratic terms in \eqref{eq:ls} and \eqref{eq:ls2}. It is well known that nonconvex functions may exhibit many stationary points, and minimizing nonconvex objectives is in general \emph{NP-hard}, and hence computationally intractable~\cite{hardproblems}.  
It is worth stressing that it is \emph{NP-hard} to establish convergence to a local minimum 
due to the existence of complicated saddle point structures~\cite{hardproblems,escaping2015ge}. 

Past approaches for solving quadratic equations can be grouped in two categories: convex and nonconvex ones. The nonconvex ones include
the `workhorse' alternating projection algorithms commonly employed in practice such as Gerchberg-Saxton~\cite{gerchberg} and Fineup~\cite{1982fienup}, AltMinPhase~\cite{altmin} and~TAF~\cite{nips2016wg,taf}, trust-region~\cite{sun2016} and Gauss-Newton methods~\cite{arxiv2016gx}, 
as well as the recently proposed Wirtinger-based variants such as (truncated/reshaped) Wirtinger flow (WF, TWF/RWF)~\cite{wf}, \cite{twf,reshaped}. 
Stochastic or incremental counterparts consisting of Kaczmarz and ITWF have been reported too~\cite{kaczmarz},~\cite{itwf}. Low-rank based phase retrieval is studied in~\cite{lrpr}. In particular, RWF can be viewed a special case of TAF without truncation, which was called simply amplitude flow in~\cite{taf} and turned out to be significantly less effective than TAF. 
On the other hand, the convex alternatives typically rely upon the so-called \emph{matrix-lifting} technique to derive semidefinite programming-based 
solvers such as PhaseLift~\cite{phaselift}, PhaseCut~\cite{phasecut}. % or solve a basis pursuit problem in the dual domain to lead to PhaseMax~\cite{phasemax}.  
Although in the context of matrix recovery, similar convex and nonconvex counterparts have also been developed  \cite{tit2016sl,2015drop,tit2015ccg,localcvx,2016pkcs,procrustes,tit2016sqw}, for concreteness the focus of this paper will be put on phase retrieval of vector signals.  
For the Gaussian model,
comparisons between convex and nonconvex solvers of problem~\eqref{eq:quad} in terms of sample size and the resulting computational complexity to acquire an $\epsilon$-accurate solution are listed in~Table~\ref{tab:comparison}.

\newpage
\begin{center}
	  \captionof{table}{Comparisons of Different Algorithms}\label{tab:comparison}\vspace{.5em}
  \begin{tabular}{ c | c | c }
    \hline
    Algorithm & Sample complexity $m$ & Computational complexity \\ \hline
  PhaseLift~\cite{phaselift}&$\mathcal{O}(n)$ &$\mathcal{O}\!\left(n^3/\epsilon^2\right)$ \\ \hline
    PhaseCut~\cite{phasecut}&$\mathcal{O}(n)$ &$\mathcal{O}\!\left(n^3/\sqrt{\epsilon}\right)$ \\ \hline
    AltMinPhase~\cite{altmin} & $\mathcal{O}\!\left(n\log n (\log^2n+\log(1/\epsilon)%\log\log(1/\epsilon))
    \right)$ & $\mathcal{O}\!\left(n^2\log n(\log^2 n+\log^2(1/\epsilon)%\log\log(1/\epsilon))
    \right)$ \\ \hline
      WF~\cite{wf}&$\mathcal{O}\!\left(n\log n\right)$ &$\mathcal{O}\!\left(n^3\log n\log(1/\epsilon)\right)$ \\ \hline
          TAF~\cite{taf}, TWF~\cite{twf}, ITWF~\cite{itwf} &$\mathcal{O}(n)$ &$\mathcal{O}\!\left(n^2\log(1/\epsilon)\right)$ \\ \hline
         %   ITWF~\cite{itwf}  &$\mathcal{O}(n)$ & $\mathcal{O}\!\left(n^2\log(1/\epsilon)\right)$ \\    \hline
    This paper&$\mathcal{O}(n)$ &$\mathcal{O}\!\left(n^2\log(1/\epsilon)\right)$ \\ \hline
  \end{tabular}
  \vspace{.5em}
\end{center}

\subsection{This Paper}

Adopting the amplitude-based nonconvex formulation, this paper puts forth a new algorithm, referred to as \emph{stochastic truncated amplitude flow} (STAF).  STAF offers an iterative solution algorithm that builds upon but considerably broadens the scope of the state-of-the-art~TAF~\cite{taf}. Specifically, it operates in two stages: Stage one employs a stochastic variance reduced gradient algorithm to obtain an orthogonality-promoting initialization, whereas the second stage applies stochastic truncated amplitude-based iterations to refine the initial estimate. Our approach is shown capable of reconstructing any $n$-dimensional real-/complex-valued signal $\bm{x}$ from a nearly minimal number of magnitude-only measurements in linear time. 
Albeit achieving the same order-optimal sample and computational complexities as TAF~\cite{taf} and~(I)TWF~\cite{twf,itwf}, STAF has two advantages. First, STAF enjoys $\mathcal{O}(n)$ iteration complexity in both initialization and refinement stages, which is order-optimal, and improves over $\mathcal{O}(n^2)$ afforded by the state-of-the-art approaches. 
Although ITWF adopts an incremental gradient method to achieve $\mathcal{O}(n)$ iteration complexity at the second stage, its first stage relies on the gradient-type power method of iteration complexity $\mathcal{O}(n^2)$ 
to obtain a truncated spectral initialization~\cite{itwf}.  
Moreover, as will be demonstrated by our simulated tests, STAF outperforms the state-of-the-art algorithms including TAF, ITWF, and (T)WF on both synthetic data and real images in terms of both exact recovery performance and convergence speed. Specifically for the real-valued Gaussian model, STAF empirically reconstructs any real-valued $n$-dimensional signal $\bm{x}$ from a number $m\approx2.3n$ of magnitude measurements, which is close to the information-theoretic limit of $m= 2n-1$. In sharp contrast, the existing alternatives such as TAF, ITWF, and (T)WF typically require a few times more 
measurements to achieve exact recovery. Markedly improved performance is also witnessed for STAF when the complex-valued Gaussian model, and coded diffraction patterns of real images~\cite{coded}, are employed.

\emph{Paper outline.} The rest of the paper is outlined as follows. 
Section~\ref{sec:alg} first reviews the truncated amplitude flow (TAF) algorithm, and subsequently motivates and derives the two stages of our proposed STAF algorithm. 
Section~\ref{sec:main} summarizes STAF, and establishes its theoretical performance. 
%The proposed TAF algorithm is summarized and its theoretical performance is established in Section~\ref{sec:main}. 
Extensive tests comparing STAF with state-of-the-art approaches on both synthetic data and real images are presented in Section~\ref{sec:test}.  
Finally, main proofs are given in Section~\ref{pf:noiseless}, and technical details can be found in the Appendix.

\section{Algorithm: Stochastic Truncated Amplitude Flow}\label{sec:alg}
In this section, TAF is first reviewed, and its limitations for large-scale applications are pointed out. 
To cope with these limitations, simple, scalable, and fast stochastic gradient descent (SGD)-type algorithms for both the initialization and gradient refinement stages are developed. 

To begin with, a number of basic concepts are introduced. Define the Euclidean distance of any estimate $\bm{z}$ to the solution set of \eqref{eq:quad} as follows: ${\rm dist}(\bm{z},\bm{x}):=
\min\left\|\bm{z}\pm\bm{x}\right\|$ for real-valued signals, and ${\rm dist}(\bm{z},\bm{x}):=\minimize_{\phi \in[0,2\pi)}\|\bm{z}-\bm{x}{\rm e}^{i\phi}\|$ for complex ones~\cite{wf}.  
Define also the indistinguishable global phase constant in the real case as
\vspace{-.em}
\begin{equation}\label{eq:adaptation}
	\phi(\bm{z}):=\left\{\begin{array}{lll}
		0,&\|\bm{z}-\bm{x}\|\le \|\bm{z}+\bm{x}\|,\\
		\pi,&{\rm otherwise.}
	\end{array}\right.
	\vspace{-.em}
\end{equation}
Henceforth, letting $\bm{x}$ be any solution of the given system in~\eqref{eq:quad}, we assume that $\phi\left({\bm{z}}\right)=0$; otherwise, ${\bm{z}}$ is replaced by ${\rm e}^{-j\phi\left({\bm{z}}\right)}{\bm{z}}$, but for brevity of exposition, the phase adaptation term ${\rm e}^{-j\phi\left({\bm{z}}\right)}$ shall be dropped whenever it is clear from the context.

\subsection{Truncated Amplitude Flow}
\label{subsec:taf}

In this section, the two stages of TAF are outlined~\cite{taf}. In stage one, TAF employs power iterations to compute an orthogonality-promoting initialization, while the second stage refines the initialization via gradient-type iterations. The orthogonality-promoting initialization 
builds upon a basic characteristic of high-dimensional spaces, which asserts that high-dimensional random vectors
 are almost always nearly orthogonal to each other~\cite{taf}. 
Its core idea relies on approximating the unknown $\bm{x}$ by a vector $\bm{z}_0\in\mathbb{R}^n$  
most orthogonal to a carefully selected subset of design vectors $\{\bm{a}_i\}_{i\in\mathcal{I}_0}$, with the index set $\mathcal{I}_0\subseteq[m]:=\{1,\,2,\,\ldots,\,m\}$. 
 It is well known that the geometric relationship between any nonzero vectors $\bm{p}\in\mathbb{R}^n$ and $\bm{q}\in\mathbb{R}^n$ can be captured by their squared normalized inner-product defined as $\cos^2\theta:={|\langle\bm{p},\bm{q}\rangle|^2}/({\|\bm{p}_i\|^2\|\bm{q}\|^2})$, where $\theta\in[0,\pi]$ signifies the angle between $\bm{p}$ and $\bm{q}$. Intuitively, the smaller $\cos^2\theta$ is, the more orthogonal the two vectors are. Assume with no loss of generality that $\|\bm{x}\|=1$, which will be justified shortly. 
Upon obtaining 
the squared normalized inner-products for all pairs $\{(\bm{a}_i,\bm{x})\}_{i=1}^m$, collectively denoted by $\{\cos^2\theta_i\}_{i=1}^m$ with $\theta_i$ denoting the angle between $\bm{a}_i$ and $\bm{x}$, 
the orthogonality-promoting initialization constructs $\mathcal{I}_0$ by including the indices of $\bm{a}_i$'s that produce one of the smallest $|\mathcal{I}_0|$ normalized inner-products. Precisely, $\bm{z}_0$ can be found by solving~\cite{taf} 
\begin{equation}
\label{eq:mineig}
	%\bm{z}_0:=\arg
	\underset{\|\bm{z}\|=1}{\text{minimize}}
~~\bm{z}^\ccalT\left(\frac{1}{|\mathcal{I}_0|}\sum_{i\in\mathcal{I}_0}\frac{\bm{a}_i\bm{a}_i^\ccalT}{\|\bm{a}_i\|^2}\right)\bm{z}
	%\\{\rm s.t.}~&~\|\bm{z}\|=1%\label{eq:mineig2}
\end{equation}	
where $|\mathcal{I}_0|$ is on the order of $n$. To be precise, as shown in \cite[Theorem 1]{taf}, one requires for exact recovery of TAF that $m\ge c_1|\mathcal{I}_0|\ge c_2 n$ holds for certain numerical constants $c_1,\,c_2>0$. Solving~\eqref{eq:mineig} amounts to finding the smallest eigenvalue and the associated eigenvector of  
$\bm{Y}_0:=\frac{1}{|\mathcal{I}_0|}\sum_{i\in\mathcal{I}_0}\frac{\bm{a}_i\bm{a}_i^\ccalT}{\|\bm{a}_i\|^2}\succeq\bm{0}$. Nevertheless, to avoid the $\mathcal{O}(n^3)$ computational complexity of computing the eigenvector associated with the smallest eigenvalue in~\eqref{eq:mineig}, an application of the standard concentration result 
$	\sum_{i=1}^m\frac{\bm{a}_i\bm{a}_i^\ccalT}{\|\bm{a}_i\|^2}
\approxeq \frac{m}{n}\bm{I}_n
$ simplifies that 
to computing the principal eigenvector of $\widebar{\bm{Y}}_0:=\frac{1}{|\widebar{\mathcal{I}}_0|}\sum_{i\in\widebar{\mathcal{I}}_0}\frac{\bm{a}_i\bm{a}_i^\ccalT}{\|\bm{a}_i\|^2}
$, where $\widebar{\mathcal{I}}_0$ is the complement of $\mathcal{I}_0$ in $[m]$. Upon collecting $\{\bm{a}_i\}_{i\in\widebar{\mathcal I}_0}$ into an $n\times |\widebar{\mathcal{I}}_0|$ data matrix $\bm{D}$, one can rewrite $\widebar{\bm Y}_0=\bm{D}\bm{D}^\ccalT$ to arrive at the following principal component analysis (PCA) problem 
\begin{equation}
	\label{eq:maxeig}
	\tilde{\bm{z}}_0:=\arg\max_{\|\bm{z}\|=1}~~\frac{1}{|\widebar{\mathcal{I}}_0|}\bm{z}^\ccalT\bm{D}\bm{D}^\ccalT
	\bm{z}.	%\widebar{\bm{Y}}_0
\end{equation} 
On the other hand, if $\|\bm{x}\|\ne 1$,  the estimate $\tilde{\bm{z}}_0$ is scaled by $\sqrt{\frac{1}{m}\sum_{i=1}^m y_i}$, a norm estimate of $\bm{x}$ to yield $\bm{z}_0:=\sqrt{\frac{1}{m}\sum_{i=1}^m y_i}\tilde{\bm z}_0$. 
Further details can be found in~\cite[Section \RN{2}.B]{taf}. 

When the signal dimension $n$ is modest, problem~\eqref{eq:maxeig} can be solved exactly by a full singular value decomposition (SVD) of $\bm{D}$~\cite{book2012matrix}. Yet it has running time of $\mathcal{O}(\min\{n^2\widebar{\mathcal{I}}_0,n\widebar{\mathcal{I}}_0^2\})$ (or simply $\mathcal{O}(n^3)$ because $|\widebar{\mathcal{I}}_0|$ is required to be on the order of $n$), which grows prohibitively in large-scale applications. 
A common alternative is the power method that is tabulated in Algorithm~\ref{alg:pm}, and was also employed by~\cite{taf,wf,twf,itwf} to find an initialization~\cite{book2012matrix}. Power method, on the other hand, involves a matrix-vector multiplication $\widebar{\bm Y}_0\bm{u}_t$  
per iteration, thus incurring iteration complexity of $\mathcal{O}(n|\widebar{\mathcal{I}}_0|)$ or $\mathcal{O}(n^2)$ by passing through the selected data $\{\bm{a}_i\}_{i\in\widebar{\mathcal{I}}_0}$.
Furthermore, to
 produce an $\epsilon$-accurate solution, it incurs runtime of~\cite{book2012matrix} 
 \begin{equation}\label{eq:pmruntime}
 \mathcal{O}\left(\frac{1}{\delta} n|\widebar{\mathcal{I}}_0|\log(1/\epsilon)\right)
 \end{equation}
depending on the eigengap 
 $\delta>0$, which is defined as the gap between the largest and the second largest eigenvalues of $\widebar{\bm Y}_0$ normalized by the largest one~\cite{book2012matrix}. It is clear that when the eigengap $\delta$ is small, the runtime of $\mathcal{O}\big(n|\widebar{\mathcal{I}}_0|\log(1/\epsilon)/\delta
 \big)$ required by the power method would be equivalent to many passes over the entire data, and this could be prohibitively for large datasets~\cite{icml2016shamir}. Hence, the power method may not be appropriate for computing the initialization in large-scale applications, particularly those involving small eigengaps.

\begin{algorithm}[h!]
  \caption{Power method}
  \label{alg:pm}
  \begin{algorithmic}[1]
\STATE {\bfseries Input:}
Matrix $\widebar{\bm Y}_0=\bm{D}\bm{D}^\ccalT$.
\STATE {\bfseries Initialize} a unit vector
 $\bm{u}_0\in\mathbb{R}^n$ randomly.
  \STATE {\bfseries For} {$t=0$ {\bfseries to} $T-1$} {\bfseries do}\\
\quad ${\bm u}_{t+1}=\frac{\widebar{\bm Y}_0{\bm u}_{t}}
{\|\widebar{\bm Y}_0{\bm u}_{t}\|}$.
\STATE {\bfseries End for}
     \STATE {\bfseries Output:}
$\tilde{\bm{z}}_0=\bm{u}_{T}$.
  \end{algorithmic}
   \vspace{-.em} 
\end{algorithm} 
  %\footnotetext{The symbol $\lceil\rceil$ is the ceiling operation returning the smallest integer greater than or equal to the given number.} 
  
  The second stage of TAF relies on truncated gradient iterations of the amplitude-based cost function~\eqref{eq:ls2}. % to refine the obtained initial estimate. 
  Specifically, with $t\ge 0$ denoting the iteration number, the truncated gradient stage starts with the initial estimate $\bm{z}_0$, and operates in the following iterative fashion
      \begin{equation}\label{eq:iter}
	\bm{z}_{t+1}=\bm{z}_t-\frac{\mu}{m}\sum_{i\in\mathcal{I}_{t+1}}\left(\bm{a}_i^\ccalT\bm{z}_t-\psi_i\frac{\bm{a}_i^\ccalT\bm{z}_t}{|\bm{a}_i^\ccalT\bm{z}_t|}
\right)\bm{a}_i  ,\quad t= 0,\,1,\,\ldots
\end{equation}
where the index set responsible for the gradient regularization is given as~\cite{taf}
\begin{equation}\label{eq:large}
	\mathcal{I}_{t+1}:=\left\{1\le i\le m\left|\frac{|\bm{a}_i^\ccalT\bm{z}_t|}{|\bm{a}_i^\ccalT\bm{x}|}\ge \frac{1}{1+\gamma}\right. \right\},\quad t= 0,\,1,\,\ldots.
\end{equation}
%The truncation indeed amounts to including only those components having $\bm{z}_t$ sufficiently away from the sign-change hyperplane $\bm{a}_i^\ccalT\bm{z}_t=0$, %i.e., $|\bm{a}_i^\ccalT\bm{z}_t|\ge \frac{1}{1+\gamma}|\bm{a}_i^\ccalT\bm{x}|$,   
%	which provably eliminates the erroneously estimated signs ($\frac{\bm{a}_i^\ccalT\bm{z}_t}{|\bm{a}_i^\ccalT\bm{z}_t|}\ne \frac{\bm{a}_i^\ccalT\bm{x}}{|\bm{a}_i^\ccalT\bm{x}|}$) with high probability. Its extreme power in the real-valued Gaussian model has been well documented in~\cite{taf}. 

\begin{figure*}[ht]
	\centering
	\begin{subfigure}
	\centering
	\includegraphics[width=.50\textwidth]{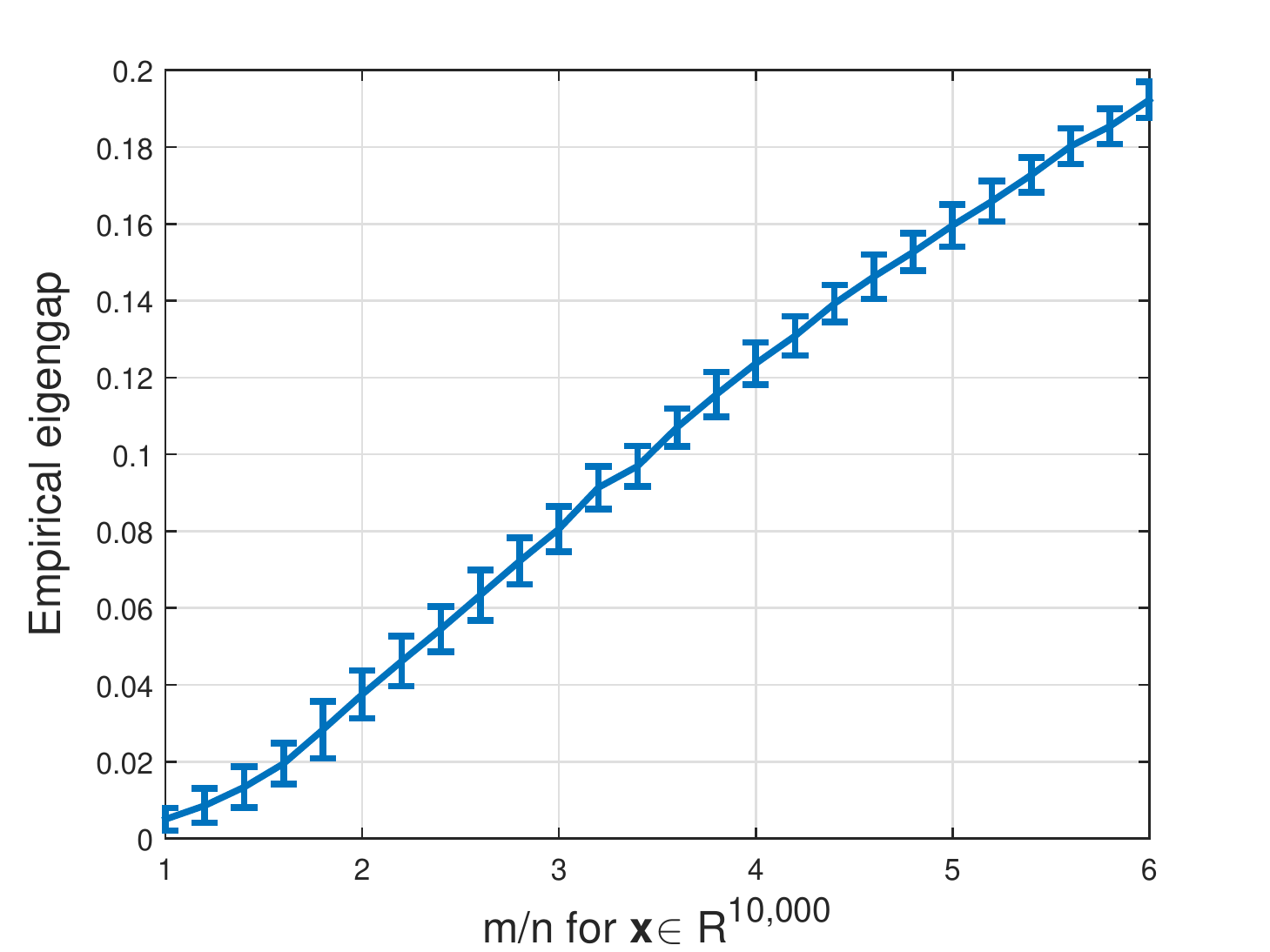}
	\end{subfigure}
	\hspace{-15pt}
	\begin{subfigure}
	\centering
	\includegraphics[width=.50\textwidth]{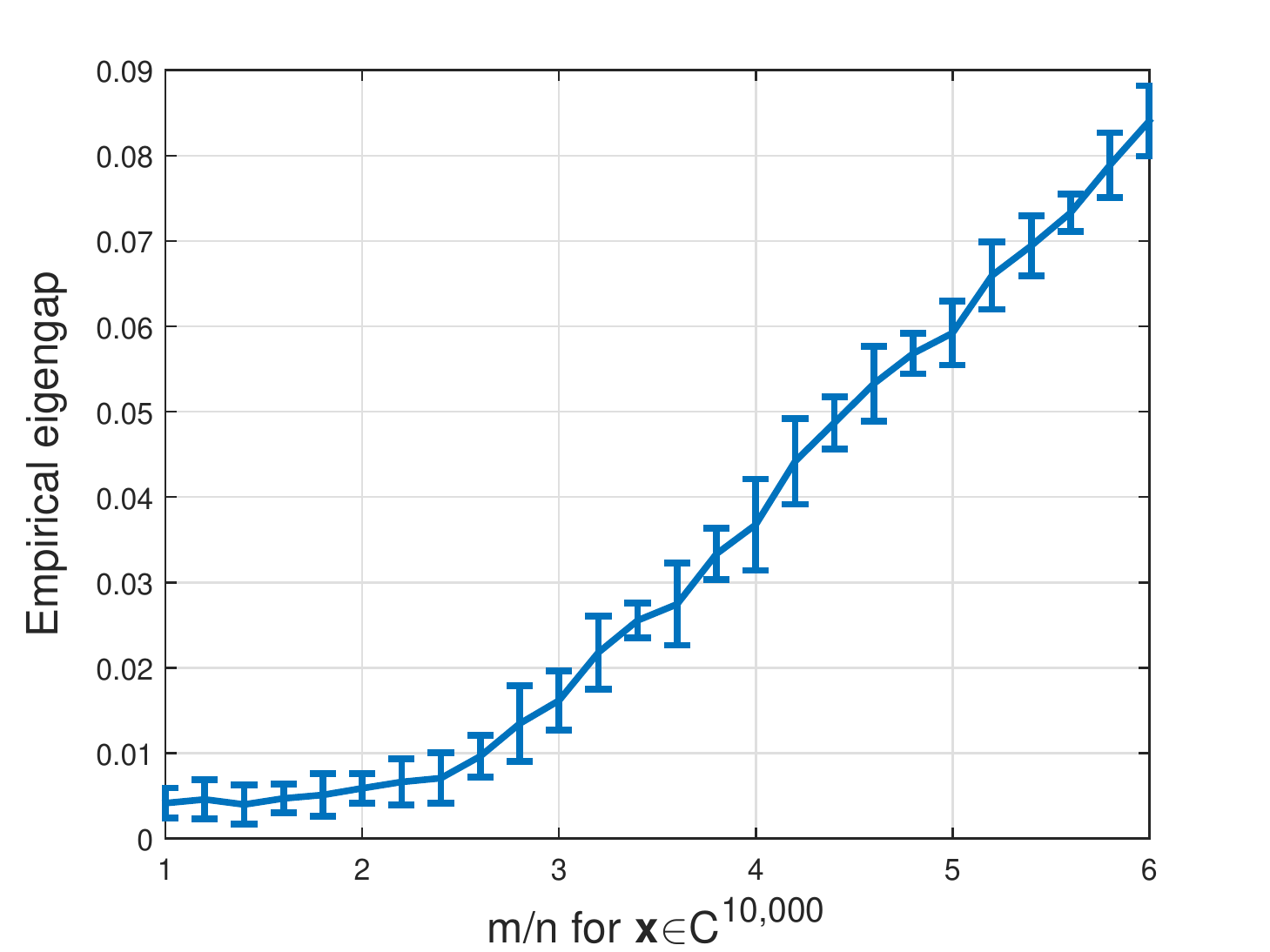}
	\end{subfigure}
    \caption{Rigengaps $\delta$ of $\bar{\bm{Y}}_0 \in \mathbb{R}^{n\times n}$ averaging over $100$ Monte Carlo realizations for $n=10,000$ fixed and $m/n$ varying by $0.2$ from $1$ to $6$. Left: Real-valued Gaussian model with $\bm{x}\sim\mathcal{N}(\bm{0},\bm{I}_n)$, and $\bm{a}_i\sim\mathcal{N}(\bm{0},\bm{I}_n)$. Right: Complex-valued Gaussian model with $\bm{x}\sim \mathcal{CN}(\bm{0},\bm{I}_n)$, and $\bm{a}_i\sim\mathcal{CN}(\bm{0},\bm{I}_n)$. }
  \label{fig:eiggap}
	\hspace{-0pt}
\end{figure*}

\subsection{Variance-reducing Orthogonality-promoting Initialization}
This section first presents some empirical evidence showing that small eigengaps appear commonly in the orthogonality-promoting initialization approach. %For PCA problems of small eigengaps, a variance-reducing stochastic alternative of better parameters in computational complexity was introduced~\cite{icml2016shamir}. 
Figure \ref{fig:eiggap} plots empirical eigengaps of $\widebar{\bm{Y}}_0\in\mathbb{R}^{n\times n}$ under the real- and complex-valued Gaussian models 
over $100$ Monte Carlo realizations under default parameters of TAF, where $n=10,000$ is fixed, and $m/n$ the number of equations and unknowns increases by $0.2$ from $1$ to $6$.  
As shown in Fig.~\ref{fig:eiggap}, the eigengaps of $\widebar{\bm{Y}}_0$ resulting from the orthogonality-promoting initialization in~\cite[Algorithm 1]{taf}
are rather small particularly for small $m/n$ close to the information limit $2$. 
 Using power iterations in Algorithm~\ref{alg:pm} 
 of runtime $\mathcal{O}\big(n|\widebar{\mathcal{I}}_0|\log(1/\epsilon)/\delta\big)$ in~\eqref{eq:pmruntime} thus entails many passes over the entire data due to a small eigengap factor of $1/\delta$, which may not perform well in the presence of large dimensions that are common to imaging applications~\cite{icml2016shamir}.    
On the other hand, instead of using the deterministic power method,  
stochastic and incremental algorithms have been advocated in~\cite{1982ojas,icml2016shamir}. These algorithms perform a much cheaper update per iteration by choosing some $i_t\in\widebar{\mathcal{I}}_0$ either uniformly at random or in a cyclic manner, and updating the current iterate using only $\bm{a}_{i_t}$. They are shown to have  iteration complexity of $\mathcal{O}(n)$, which is very appealing to large-scale applications.  Building on recent advances in accelerating stochastic optimization schemes~\cite{svrg2013}, a  
 variance-reducing principal component analysis (VR-PCA) algorithm was proposed in~\cite{icml2016shamir}.  VR-PCA performs cheap stochastic iterations, yet its total runtime is $\mathcal{O}\big(n(\widebar{\mathcal{I}}_0+1/\delta^2)\log(1/\epsilon)\big)$ which depends only logarithmically on the solution accuracy $\epsilon>0$. 
 This is in sharp contrast to the standard SGD variant, % of sublinear convergence rate, 
 whose runtime depends on $1/\epsilon$~\cite{1982ojas}. For the considered large-scale phase retrieval in most imaging applications, this paper advocates using the state-of-the-art algorithm VR-PCA to solve the orthogonality-promoting initialization problem in~\eqref{eq:maxeig}. In this direction, 
we refer to the resulting solution algorithm at the initialization stage as the \emph{variance-reducing orthogonality-promoting initialization} (VR-OPI), which is summarized in Algorithm~\ref{alg:vropi} next.
Specifically, VR-OPI is a double-looped algorithm with a single execution of the inner loop referred to as an iteration and one execution of the outer loop referred to as an epoch. In practice, the algorithm consists of $S$ epochs, while each epoch runs $T$ (typically taken to be the data size $|\widebar{\mathcal{I}}_0|$) iterations. 

\begin{algorithm}[h!]
	\caption{Variance-reduced orthogonality-promoting initialization (VR-OPI)}
	\label{alg:vropi}
	\begin{algorithmic}[1]
		\STATE {\bfseries Input:}
		Data matrix ${\bm D}=\{\bm{a}_i\}_{i\in\widebar{\mathcal{I}}_0}$, step size $\eta=20/m$, as well as the number of epochs $S=100$, and the epoch length $T=|\widebar{\mathcal{I}}_0|$ (by default).
		\STATE {\bfseries Initialize} a unit vector
		$\tilde{\bm{u}}_0\in\mathbb{R}^n$ randomly.
		\STATE {\bfseries For $s=0$ {\bfseries to} $S-1$ do}
			\\
		\quad	$\bm{w}=\frac{1	}{|\widebar{\mathcal{ I}}_0|}\sum_{i\in\widebar{\mathcal{I}}_0}\bm{a}_i(\bm{a}_i^\ccalT\tilde{\bm{u}}_{s})$\\
		\quad $\bm{u}_1=\tilde{\bm{u}}_{s}$.
				\STATE {\bfseries \quad  For} {$t=0$ {\bfseries to} $T-1$ do}\\
				\quad\quad Pick $i_t\in\widebar{\mathcal{I}}_0$ uniformly at random\\
				\quad\quad ${\bm \nu}_{t+1}=\bm{u}_t+\eta\left[\bm{a}_{i_t}\!\left(\bm{a}_{i_t}^\ccalT\bm{u}_t-\bm{a}_{i_t}^\ccalT\tilde{\bm{u}}_{s}\right)+\bm{w}\right]$\\
				\quad\quad $\bm{u}_{t+1}=\frac{{\bm \nu}_{t+1}}{\|{\bm \nu}_{t+1}\|}$.\\
				\STATE{\bfseries \quad End For}\\
				\quad $\tilde{\bm{u}}_{s+1}=\bm{u}_T$.
						\STATE{\bfseries  End For}\\
		\STATE {\bfseries Output:}
								$\tilde{\bm{z}}_0=\bm{u}_{S}$.
	\end{algorithmic}
	\vspace{-.em} 
\end{algorithm}

%Although effecting simple stochastic iterations, VR-OPI still enjoys a linear convergence rate under standard conditions. 
The following results adopted from~\cite[Theorem 1]{icml2016shamir} establish linear convergence rate of VR-OPI. 
\begin{proposition}[\cite{icml2016shamir}]\label{prop:vropi}
	Let $\bm{v}_1\in\mathbb{R}^n$ be an eigenvector of $\widebar{\bm{Y}}_0$ associated with the largest eigenvalue $\lambda_1$. Assume that $\max_{i\in[m]}\|\bm{a}_i\|^2\le r:=2.3n$ (which holds with probability at least $ 1-m{\rm e}^{-n/2}$), the two largest eigenvalues of $\widebar{\bm Y}_0$ are $\lambda_1>\lambda_2>0$ with eigengap $\delta=(\lambda_1-\lambda_2)/\lambda_1$, and that $\langle\tilde{\bm{u}}_0,\bm{v}_1\rangle\ge 1/\sqrt{2}$. With any $0<\epsilon,\,\xi<1$, constant step size $\eta>0$, and epoch length $T$ chosen such that
	\begin{equation}\label{eq:consts}
		\eta\le \frac{c_0\xi^2}{r^2}\delta,\quad T\ge \frac{c_1\log(2/\xi)}{\eta\delta}, \quad T\eta^2r^2+r\eta\sqrt{T\log(2/\xi)}\le c_2
	\end{equation} 
for certain universal constants $c_0,\,c_1,\,c_2>0$, successive estimates of VR-OPI (summarized in Algorithm~\ref{alg:vropi}) after $S=\left\lceil{\log(1/\epsilon)}/{\log(2/\xi)}\right\rceil$ epochs 
satisfy 
\begin{equation}\label{eq:epsilon}
	\left|\left\langle\tilde{\bm u}_S,\bm{v}_1\right\rangle\right|^2\ge 1-\epsilon
\end{equation}
with probability exceeding $1-\lceil{\log\epsilon}\rceil\xi$. Typical parameter values are $\eta=20/m$, $S=100$, and $T=|\widebar{ \mathcal{I}}_0|$.  
\end{proposition}

The proof of Proposition~\ref{prop:vropi} can be adapted from~\cite{icml2016shamir}. Even though PCA in~\eqref{eq:maxeig} is nonconvex, the SGD based VR-OPI algorithm converges 
 to the globally optimal solution under mild conditions~\cite{icml2016shamir}.   
Moreover, fixing any $\xi\in(0,1)$, conditions in~\eqref{eq:consts} hold true when $T$ is chosen to be on the order of $1/(\eta\delta)$, and $\eta$ to be sufficiently smaller than $\delta/r^2$. Expressed differently, if VR-OPI runs $T=\Theta(r^2/\delta^2)$
 iterations per epoch for a total number $S=\Theta\big(\log(1/\epsilon)\big)$ 
of epochs, then the returned VR-OPI estimate is $\epsilon$-accurate with probability at least $1-\lceil\log_2(1/\epsilon)\rceil\xi$. Since each epoch takes $\mathcal{O}\big( n(T+|\widebar{\mathcal{I}}_0|)\big)$ time to implement, the total runtime is of
\begin{equation}
	\label{eq:runtime}
	\mathcal{O}\!\Big(n\Big(|\widebar{\mathcal{I}}_0|+\frac{r^2}{\delta^2}
	\Big)\log(1/\epsilon)
	\Big)
\end{equation} 
which validates the exponential convergence rate of VR-OPI. In addition, when $\delta/r\ge \Omega\big(1\big/\sqrt{|\widebar{\mathcal{I}}_0|}\big)$, 
the total runtime reduces to $\mathcal{O}\big(n|\widebar{\mathcal{I}}_0|\log(1/\epsilon)\big)$ up to log-factors. It is worth emphasizing that 
the required runtime is proportional to the time required to scan the selected data once, which is in stark contrast to the runtime of $\mathcal{O}\big(n|\widebar{\mathcal{I}}_0|\log(1/\epsilon)/\delta\big)$ when using power method~\cite{book2012matrix}. Simulated tests in Section~\ref{sec:test} corroborate the effectiveness of VR-OPI over the popular power method in processing data involving large dimensions $m$ and/or $n$.

\subsection{Stochastic Truncated Gradient Stage}
Driven by the need of efficiently processing
large-scale phaseless data in imaging applications, a stochastic solution algorithm is put forth for minimizing the amplitude-based cost function in~\eqref{eq:ls2}. To ensure good performance, the gradient regularization rule in~\eqref{eq:large} is also accounted for to lead to our truncated stochastic gradient iterations. 
%Interestingly though, the utilization of the gradient regularization rule also markedly speeds up the algorithm's convergence relative to its `plain-vallina' version, which will be validated through the simulated tests in Section~\ref{sec:test}.    
It is worth mentioning that the Kaczmarz method~\cite{1937kaczmarz} was also used for solving a system of phaseless quadratic equations
in~\cite{kaczmarz}. 
However, Kaczmarz variants of block or randomized updates converge to at most a neighborhood of the optimal solution $\bm{x}$. Distance between the Kaczmarz estimates and $\bm{x}$ is bounded in terms of the dimension $m$ and the size of the amplitude data vector $\bm{\psi}$ measured by the $\ell_1$- or $\ell_\infty$-norm. Nevertheless, the obtained bounds of the form $m\|\bm{\psi}\|_{1}$ or $m\|\bm{\psi}\|_{\infty}$ are rather loose ($m$ typically very large), and less attractive than the geometric convergence to the global solution $\bm{x}$ to be established for also stochastic iterations based STAF.    

Adopting the intensity-based Poisson likelihood function~\eqref{eq:ls}, an incremental version of TWF was developed in~\cite{itwf}, which provably converges to $\bm{x}$ in linear time. 
Albeit achieving improved empirical performance and faster convergence over TWF in terms of the number of passes over the entire data to produce an $\epsilon$-accurate solution~\cite{twf}, the number of measurements it requires for exact recovery is still relatively far from the information-theoretic limits. Specifically for the real-valued Gaussian $\bm{a}_i$ designs, ITWF requires about $m\ge 3.2n$ noiseless measurements to guarantee exact recovery relative to $4.5n$ for TWF~\cite{twf}.
 Recall that TAF achieves exact recovery from about $3n$ measurements~\cite{taf}. Furthermore, gradient iterations can be easily trapped in saddle points when dealing with nonconvex optimization. In contrast, stochastic iterations are able to escape saddle points, and converge globally to at least a local minimum~\cite{escaping2015ge}. 
Hence, besides the appealing computational advantage, 
  stochastic counterparts of TAF may further improve the performance over TAF, as also asserted by the comparison between ITWF and TWF.
In the following, we present two STAF variants: Starting with an initial estimate $\bm{z}_0$ found
using VR-OPI in Algorithm~\ref{alg:vropi},  
the first variant successively updates $\bm{z}_0$ through amplitude-based stochastic gradient iterations with a constant step size $\mu>0$ chosen on the order of $1/n$,  
while the second operates much like the  Kaczmarz method, 
yet both suitably account for the truncation rule in~\eqref{eq:large}. To understand stochastic gradient descent and the Kaczmarz method and also their connections, interested readers can refer to~\cite{nips2014nws}.

For simplicity of exposition, let us rewrite the amplitude-based cost function as follows
\begin{equation}\label{eq:lscost}
\underset{\bm{z}\in\mathbb{R}^n}{\text{minimize}}~~\ell(\bm{z})=\sum_{i=1}^m\ell_i(\bm{z})
%=\|\bm{\psi}-|\bm{a}_i^\ccalT\bm{z}|\|^2
:=\frac{1}{2}\sum_{i=1}^m\left(\psi_i-|\bm{a}_i^\ccalT\bm{z}|\right)^2
\end{equation}
where the factor $1/2$ is introduced for notational convenience. 
%where $|\bm{A}\bm{z}|:=\big[|\bm{a}_i^\ccalT\bm{z}|~s~|\bm{a}_i^\ccalT\bm{z}|\big]^\ccalT$ with a slight abuse of notation. 
It is clear that the cost $\ell(\bm{z})$ or each $\ell_i(\bm{z})$ in~\eqref{eq:lscost} is nonconvex and nonsmooth; hence, the optimization  in \eqref{eq:lscost} is
computationally intractable in general~\cite{nphard}. %It is known that the generalized gradient of nonconvex nonsmooth functions plays the role of the subgradient in convex nonsmooth optimization~\cite{shor2012}. 
Along the lines of nonconvex paradigms including WF~\cite{wf} and TAF~\cite{taf}, our approach to solving the problem at hand 
amounts to iteratively refining the initial estimate $\bm{z}_0$ 
by means of truncated stochastic gradient iterations. This is in contrast to (T)WF and TAF, which rely on (truncated) gradient-type iterations~\cite{wf,twf,taf}. STAF processes one datum at a time and evaluates the generalized gradient of one component function $\ell_{i_t}(\bm{z})$ for some index $i_t\in \{1,2,\ldots,m\}$ per iteration $t\ge 0$. Specifically, STAF successively updates $\bm{z}_0$ using   
 the following truncated stochastic gradient iterations
\begin{equation}
	\bm{z}_{t+1}=\bm{z}_t-\mu_t\nabla \ell_{i_t}(\bm{z}_t) \mathbb{1}_{\{|\bm{a}_{i_t}^\ccalT\bm{z}_t|/|\bm{a}_{i_t}^\ccalT\bm{x}|\ge 1/(1+\gamma)\}},\quad t=0,\,1,\,\ldots\label{eq:stafiter}
\end{equation}
with 
\begin{equation}
	\nabla \ell_{i_t}(\bm{z}_t)=\Big(\bm{a}_{i_t}^\ccalT\bm{z}_t-\psi_{i_t}\frac{\bm{a}_{i_t}^\ccalT\bm{z}_t}{|\bm{a}_{i_t}^\ccalT\bm{z}_t|}\Big)\bm{a}_{i_t}
\end{equation}
where $\mu_t$ is either set to be a constant $\mu>0$ on the order of $1/n$, or taken as the time-varying one as in Kaczmarz's iteration, namely, $\mu_t=1/\|\bm{a}_{i_t}\|^2$~\cite{1937kaczmarz}. The index $i_t$ is sampled uniformly at random or with given probabilities from $\{1,\,2,\ldots,m\}$, or it simply cycles through the entire set $[m]$. 
%The term $\partial \ell_{i_t}(\bm{z}_t)$ is a generalized gradient of the nonconvex and nonsmooth function $\ell_{i_t}(\bm{z})$ evaluated at $\bm{z}_t$, which corresponds to a subgradient in the convex nonsmooth optimization regime. More details regarding the generalized gradient can be found in \cite[Definition 1]{clarke1975gg}, \cite{taf}. 
In addition, fixing the truncation threshold to $\gamma=0.7$, 
the indicator function $\mathbb{1}_{\{|\bm{a}_{i_t}^\ccalT\bm{z}_t|/|\bm{a}_{i_t}^\ccalT\bm{x}|\ge 1/(1+\gamma)\}}$ in \eqref{eq:stafiter} takes the value $1$,  if $|\bm{a}_{i_t}^\ccalT\bm{z}_t|/|\bm{a}_{i_t}^\ccalT\bm{x}|\ge 1/(1+\gamma)$ holds true; and $0$ otherwise. It is worth stressing that this truncation rule provably rejects `bad' search directions with high probability. Moreover, this regularization maintains only gradient components of large enough $|\bm{a}_i^\ccalT\bm{z}^t|$ values, hence saving the objective function~\eqref{eq:ls1} from being non-differentiable at $\bm{z}^t$ and simplifying the theoretical analysis. In the context of large-scale linear regressions, similar ideas such as censoring have been advocated for speeding up SGD-type algorithms~\cite{icassp2015dbkv},~\cite{globalsip2014wang}. 
Numerical tests demonstrating the performance improvement using the stochastic truncated iterations will be presented in Section~\ref{sec:test}.

\begin{algorithm}[h!]
  \caption{Stochastic truncated amplitude flow (STAF) algorithm}
  \label{alg:STAF}
  \begin{algorithmic}[1]
\STATE {\bfseries Input:}
Sampling vectors $\{\bm{a}_i\}_{i=1}^m$, and data $\left\{\psi_i:=\left|\langle\bm{a}_i,\bm{x}\rangle\right|\right\}_{i=1}^m$;  maximum number of iterations $T=500m$; by default, constant step sizes $\mu=0.8/n$ or $\mu=1.2/n$ in the real- or complex-valued Gaussian models, truncation thresholds $|\widebar{\mathcal{I}}_0|=\lceil\frac{1}{6}m\rceil$, %\footnotemark
~and $\gamma=0.7$.
\STATE {\bfseries Evaluate} $\widebar{\mathcal{I}}_0$ to consist of indices associated with the $|\widebar{\mathcal{I}}_0|$ largest values among $\left\{\psi_i/\|\bm{a}_i\|\right\}$.
\STATE {\bfseries Initialize} $\bm{z}_0$ as $\sqrt{\frac{1}{m}\sum_{i=1}^m\psi_i^2}\tilde{\bm{z}}_0$, where $\tilde{\bm{z}}_0$ is obtained via  Algorithm~\ref{alg:vropi} with $\widebar{\bm{Y}}_0:=\frac{1}{|\widebar{\mathcal{I}}_0|}\sum_{i\in\widebar{\mathcal{I}}_0}\frac{\bm{a}_i\bm{a}_i^\ccalT}{\|\bm{a}_i\|^2}$.
  \STATE {\bfseries For} {$t=0$ {\bfseries to} $T-1$ do}\\
{ \begin{equation} \label{eq:ssgd}
   	\bm{z}_{t+1}=\bm{z}_t-\mu\bigg(\bm{a}_{i_t}^\ccalT\bm{z}_t-\psi_{i_t}\frac{\bm{a}_{i_t}^\ccalT\bm{z}_t}{|\bm{a}_{i_t}^\ccalT\bm{z}_t|}\bigg)\bm{a}_{i_t}\mathbb{1}_{\big\{\left|\bm{a}_{i_t}^\ccalT\bm{z}_t\right|\ge \frac{1}{1+\gamma}{\psi_{i_t}}\big\}}
\end{equation}
where $i_t$ is sampled uniformly at random from $\{1,2,\ldots,m\}$, 
or, 
 \begin{equation}  \label{eq:kstaf}
   	\bm{z}_{t+1}=\bm{z}_t-\frac{1}{\|\bm{a}_{i_t}\|^2}\bigg(\bm{a}_{i_t}^\ccalT\bm{z}_t-\psi_{i_t}\frac{\bm{a}_{i_t}^\ccalT\bm{z}_t}{|\bm{a}_{i_t}^\ccalT\bm{z}_t|}\bigg)\bm{a}_{i_t}\mathbb{1}_{\big\{\left|\bm{a}_{i_t}^\ccalT\bm{z}_t\right|\ge \frac{1}{1+\gamma}{\psi_{i_t}}\big\}}
\end{equation}
 where $i_t$ is sampled at random from $\{1,2,\ldots,m\}$ with probability proportional to $\|\bm{a}_{i_t}\|^2$.
 }
      \STATE {\bfseries End for}
     \STATE {\bfseries Output:}
$\bm{z}_{T}$.
  \end{algorithmic}
\end{algorithm} 
	%\footnotetext{The notation $\lceil\cdot\rceil$ is the ceiling operation that returns the smallest integer greater than or equal to the given number.} 

\section{Main Results}\label{sec:main}

The proposed STAF scheme is summarized as Algorithm~\ref{alg:STAF}, with either  constant step size $\mu>0$ in the truncated stochastic gradient iterations in~\eqref{eq:ssgd},
or with time-varying step size $\mu_t=1/\|\bm{a}_{i_t}\|^2$ in the truncated Kaczmarz iterations in~\eqref{eq:kstaf}. Equipped with an initialization obtained using VR-OPI, both STAF variants will be shown to converge at an exponential rate to the globally optimal solution with high probability, as soon as $m/n$ the number of equations and unknowns exceeds some numerical constant.  

Assuming $m$ independent data samples $\{(\bm{a}_i;\psi_i)\}$ drawn from the real-valued Gaussian model,  
the following theorem establishes  theoretical performance of STAF in the absence of noise.  
\begin{theorem}[\bf Exact recovery]\label{thm:noiseless}
	Consider the noiseless measurements $\psi_i=|\bm{a}_i^\ccalT\bm{x}|$ with an arbitrary signal $\bm{x}\in\mathbb{R}^n$, and i.i.d. $\{\bm{a}_i\sim \mathcal{N}(\bm{0},\bm{I}_n)\}_{i=1}^m$. If $\mu_t$ is either set to be a constant $\mu>0$ as per \eqref{eq:ssgd},
	or it is time-varying $\mu_t=1/\|\bm{a}_{i_t}\|^2$ as per \eqref{eq:kstaf} with the corresponding index sampling scheme, 
and also	\begin{equation}
		m\ge c_0 n\quad\text{and}\quad \mu\le \mu_0/n
	\end{equation}
	then with probability at least $1-c_1m\exp(-c_2n)$, the stochastic truncated amplitude flow (STAF) estimates (tabulated in Algorithm~\ref{alg:STAF} with default parameters) satisfy 
	\begin{equation}\label{eq:error}
		\mathbb{E}_{\mathcal{P}_t}\left[{\rm dist}^2(\bm{z}_t,\bm{x})\right]\le \rho\left(1-\frac{\nu}{n}\right)^t\|\bm{x}\|^2,\quad t=0,\,1,\,\ldots
	\end{equation} 
	for $\rho=1/10$ and some numerical constant $\nu>0$, 
	where the expectation is taken over the path sequence $\mathcal{P}_t:=\{i_0,\,i_1,\ldots,i_{t-1}\}$, and $c_0,\,c_1,\,c_2,\,\mu_0>0$ are certain universal constants. 
\end{theorem}

The proof of Theorem~\ref{thm:noiseless} is deferred to Section~\ref{pf:noiseless}.  
Apparently, the mean-square distance between the iterate and the global solution is reduced by a factor of $(1-\nu/n)^m$ after one pass through the entire data. 
Heed that the expectation $\mathbb{E}_{\mathcal{P}_t}[\cdot]$ in \eqref{eq:error} is taken over the algorithmic randomness $\mathcal{P}_t$ rather than the data. This is important since in general the data may be modeled as deterministic. 
Although only performing stochastic iterations in \eqref{eq:ssgd} and \eqref{eq:kstaf}, STAF still enjoys linear convergence rate. This is in sharp contrast to typical SGD methods, where variance reduction techniques controlling the variance of the stochastic gradients are required to achieve linear convergence rate~\cite{svrg2013,icml2016shamir}, as in~Algorithm~\ref{alg:vropi}. Moreover, the largest constant step size that STAF can afford is estimated to be $\mu_0=1.0835$, giving rise to a convergence factor of $\nu_0=0.1139$ in \eqref{eq:error}. When truncated Kaczmarz iterations are implemented, $\nu$ is estimated to be $1.5758$ much larger than the one in the constant step size case. Our experience with numerical experiments also confirm that the Kaczmarz-based STAF in \eqref{eq:kstaf} converges faster than the constant step-size based one in~\eqref{eq:ssgd}, yet it is slightly more sensitive when additive noise is present in the data.

%When noisy measurements are accounted for, the following noisy data model is considered~\cite{altmin}
%\begin{equation}
%	\label{eq:noisy}
%	\psi_i=|\bm{a}_i^\ccalT\bm{x}+\eta_i|
%\end{equation}

\section{Simulated Tests}\label{sec:test}
This section presents extensive numerical experiments evaluating the performance of STAF
using both synthetic data and real images. STAF was thoroughly compared with  existing alternatives including
TAF~\cite{taf}, (T)WF~\cite{wf,twf}, and ITWF~\cite{itwf}.
For fair comparisons, all the parameters pertinent to implementation of each algorithm were set to their suggested values.
The initialization in each scheme was found based on a number of (power/stochastic) iterations
equivalent to $100$ passes over the entire data, 
which was subsequently refined by a number of iterations corresponding to $1,000$ passes; unless otherwise stated.
All simulated estimates were averaged over $100$ independent Monte Carlo trials. 
Two performance evaluation metrics were used: the relative root mean-square error defined as
${\rm Relative~error}:={\rm dist}(\bm{z},\bm{x})/\|\bm{x}\| $;
and the empirical successful recovery rate among $100$ independent runs, in which a success is declared 
when the returned estimate incurs a relative error less than $10^{-5}$~\cite{wf}. 
Tests using both noiseless/noisy real-/complex-valued Gaussian models $\psi_i=|\bm{a}_i^\ccalH\bm{x}|+\eta_i$ 
were conducted, where the i.i.d. noise obeys $\eta_i\sim\mathcal{N}(0,\sigma^2\|\bm{x}\|^2)$. 
The Matlab implementations of STAF can be downloaded from \url{http://www.tc.umn.edu/~gangwang/STAF.}

\begin{figure}[ht]
	\centering
	\begin{subfigure}
	\centering
	\includegraphics[width=.50\textwidth]{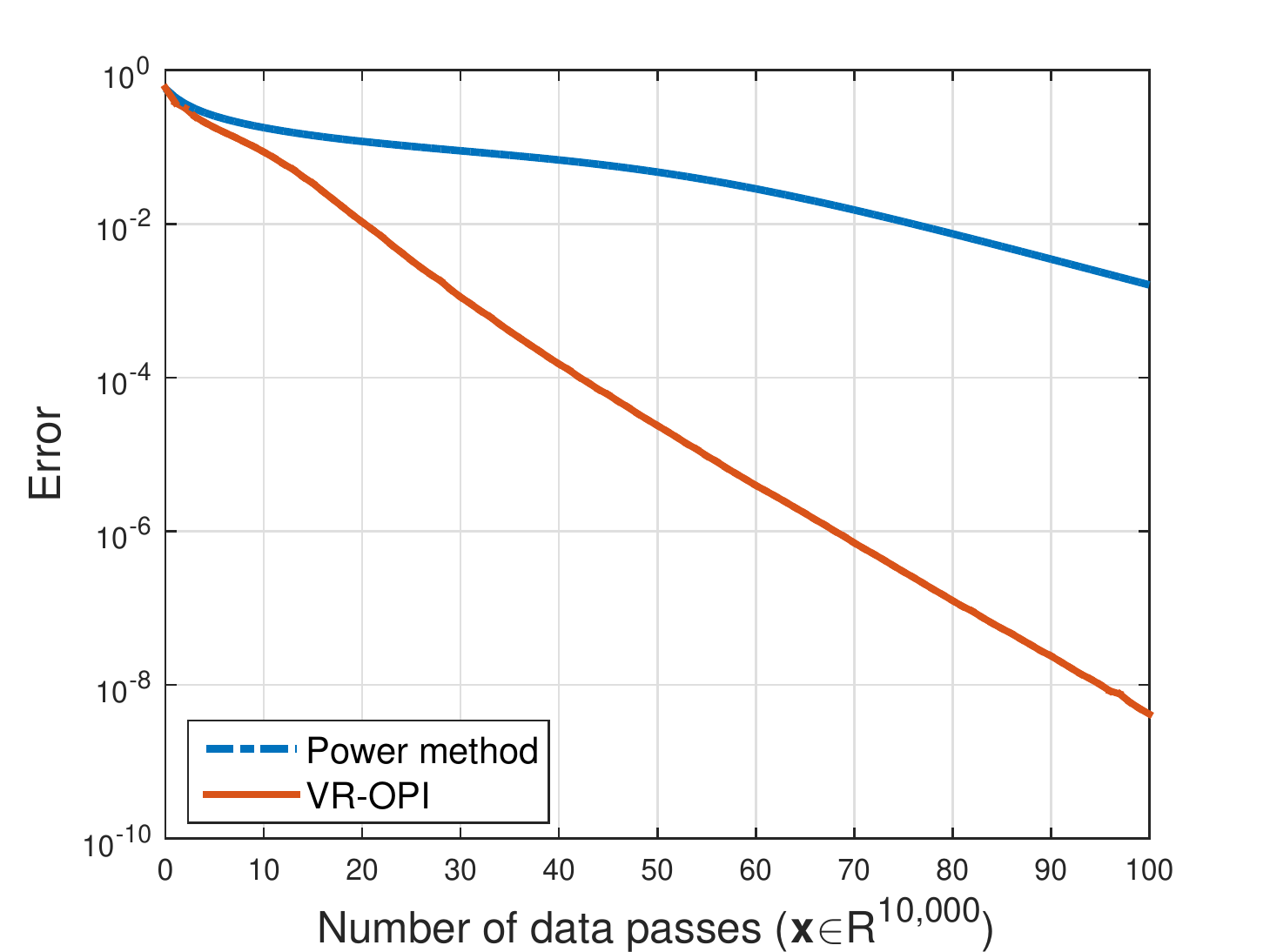}
	\end{subfigure}
	\hspace{-12pt}
	\begin{subfigure}
	\centering
	\includegraphics[width=.50\textwidth]{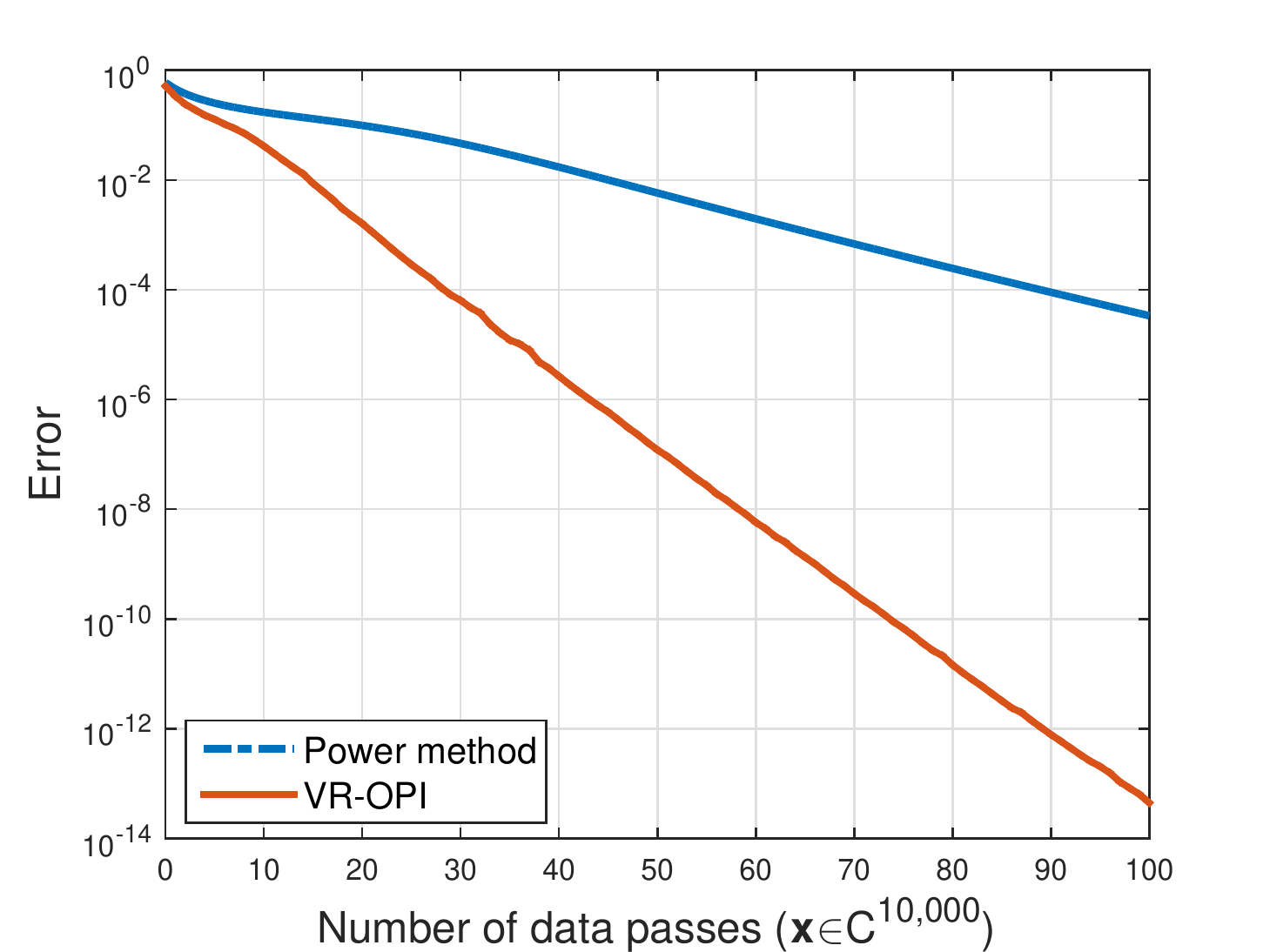} 
	\end{subfigure}
\caption{
Error evolution of the iterates
using: i) power method in Algorithm~\ref{alg:pm}; and ii) variance-reducing orthogonality-promoting initialization in Algorithm~\ref{alg:vropi} for solving problem~\eqref{eq:maxeig} with step size $\eta=1$. Left: Noiseless real-valued Gaussian model with $\bm{x}\sim\mathcal{N}(\bm{0},\bm{I}_n)$, and  $\bm{a}_i\sim\mathcal{N}(\bm{0},\bm{I}_n)$, where $n=10^4$, and $m=2n-1$. Right: Noiseless complex-valued Gaussian model with $\bm{x}\sim \mathcal{CN}(\bm{0},\bm{I}_n)$, and $\bm{a}_i\sim\mathcal{CN}(\bm{0},\bm{I}_n)$, where $n=10^4$, and $m=4n-4$. 
}
\label{fig:syninit}
	\hspace{-0pt}
\end{figure}

The first experiment compares VR-OPI in Algorithm~\ref{alg:vropi}
with the power method in Algorithm~\ref{alg:pm} 
to solve the orthogonality-promoting initialization optimization in~\eqref{eq:maxeig}. 
The comparison is carried out in terms of the number of data passes to achieve the same solution accuracy, in which one pass through the selected data amounts to a number $|\widebar{\mathcal{I}}_0|$ of gradient evaluations of component functions.
First, synthetic data based experiments are conducted using the real-/complex-valued Gaussian models
with $n=10,000$ under the known sufficient conditions for uniqueness, i.e., $m=2n-1$ in the real case, and $m=4n-4$ in the complex case. Figure~\ref{fig:syninit} plots the error evolution of the iterates $\bm{u}_t$ for the power method and VR-OPI, where
the error in logarithmic scale is defined as 
$\log_{10}\left(1-{\|\bm{D}^\ccalT\bm{u}_t\|^2}/{\|\bm{D}^\ccalT\bm{v}_0\|^2}\right)$ with the exact principal eigenvector $\bm{v}_0$ computed from the SVD of $\widebar{\bm{Y}}_0=\bm{D}\bm{D}^\ccalT$ in~\eqref{eq:maxeig}. 
Apparently, the inexpensive stochastic iterations of VR-OPI achieve certain solution accuracy with considerably fewer gradient evaluations or data passes in both real and complex settings. This is important for tasks of large $|\widebar{\mathcal{I}}_0|$, or equivalently large dimension $m$ (since $|\widebar{\mathcal{I}}_0|=5m/6$ by default), because one less data pass 
implies $|\widebar{\mathcal{I}}_0|$ fewer gradient evaluations and thus results in considerable savings in computational resources.

\begin{figure}[ht!]
	\centering
	\begin{subfigure}
	\centering
	\includegraphics[width=.50\textwidth]{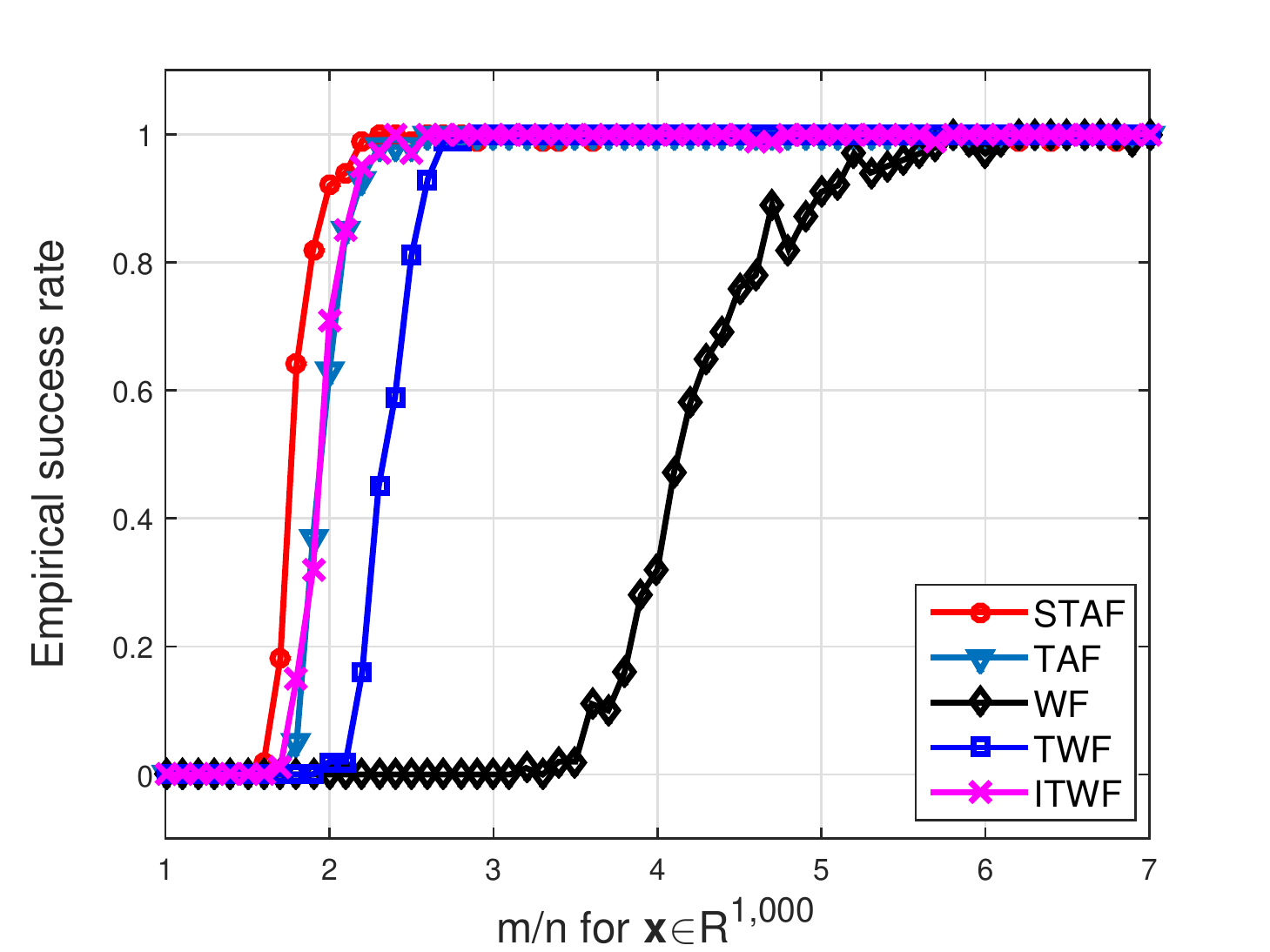}
	\end{subfigure}
	\hspace{-12pt}	
	\begin{subfigure}
	\centering
	\includegraphics[width=.50\textwidth]{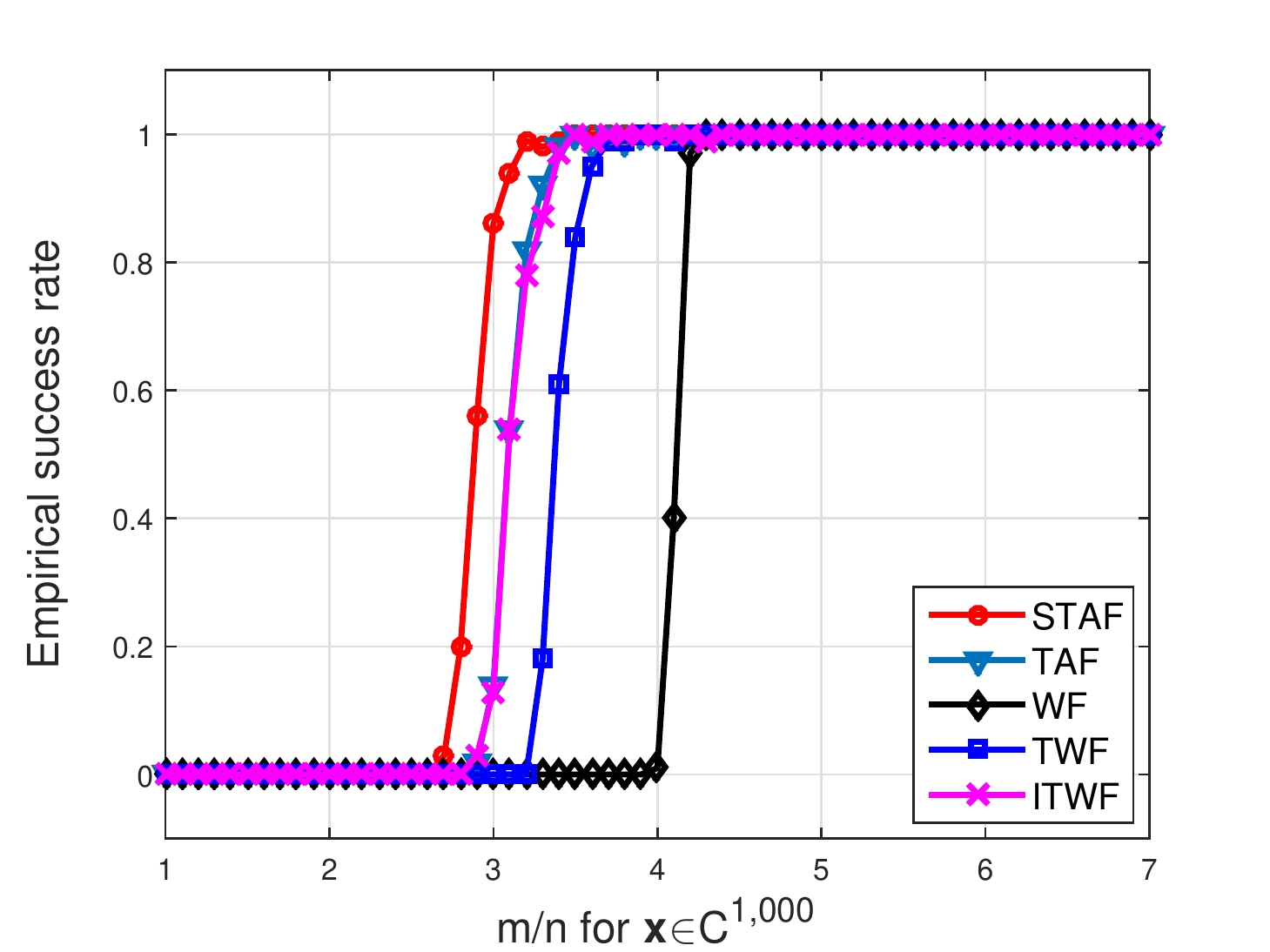} 
	\end{subfigure}
\caption{Empirical success rate for: i) WF~\cite{wf}; ii) TWF~\cite{twf}; iii)~ITAF~\cite{itwf}; iv) TAF~\cite{taf}; and v) STAF with $n=1,000$ and $m/n$ varying by $0.1$ from $1$ to $7$ under the \emph{same} orthogonality-promoting initialization. Left: Noiseless real-valued Gaussian model with 
$\bm{x}\sim\mathcal{N}(\bm{0},\bm{I}_n)$, and $\bm{a}_i\sim\mathcal{N}(\bm{0},\bm{I}_n)$; Right: Noiseless complex-valued Gaussian model with 
 $\bm{x}\sim\mathcal{CN}(\bm{0},\bm{I}_n)$, and $\bm{a}_i\sim\mathcal{CN}(\bm{0},\bm{I}_n)$.}
\label{fig:2ndrate}
\vspace{-0pt}
\end{figure}

\begin{figure}[ht!]
	\centering
	\begin{subfigure}
	\centering
	\includegraphics[width=.50\textwidth]{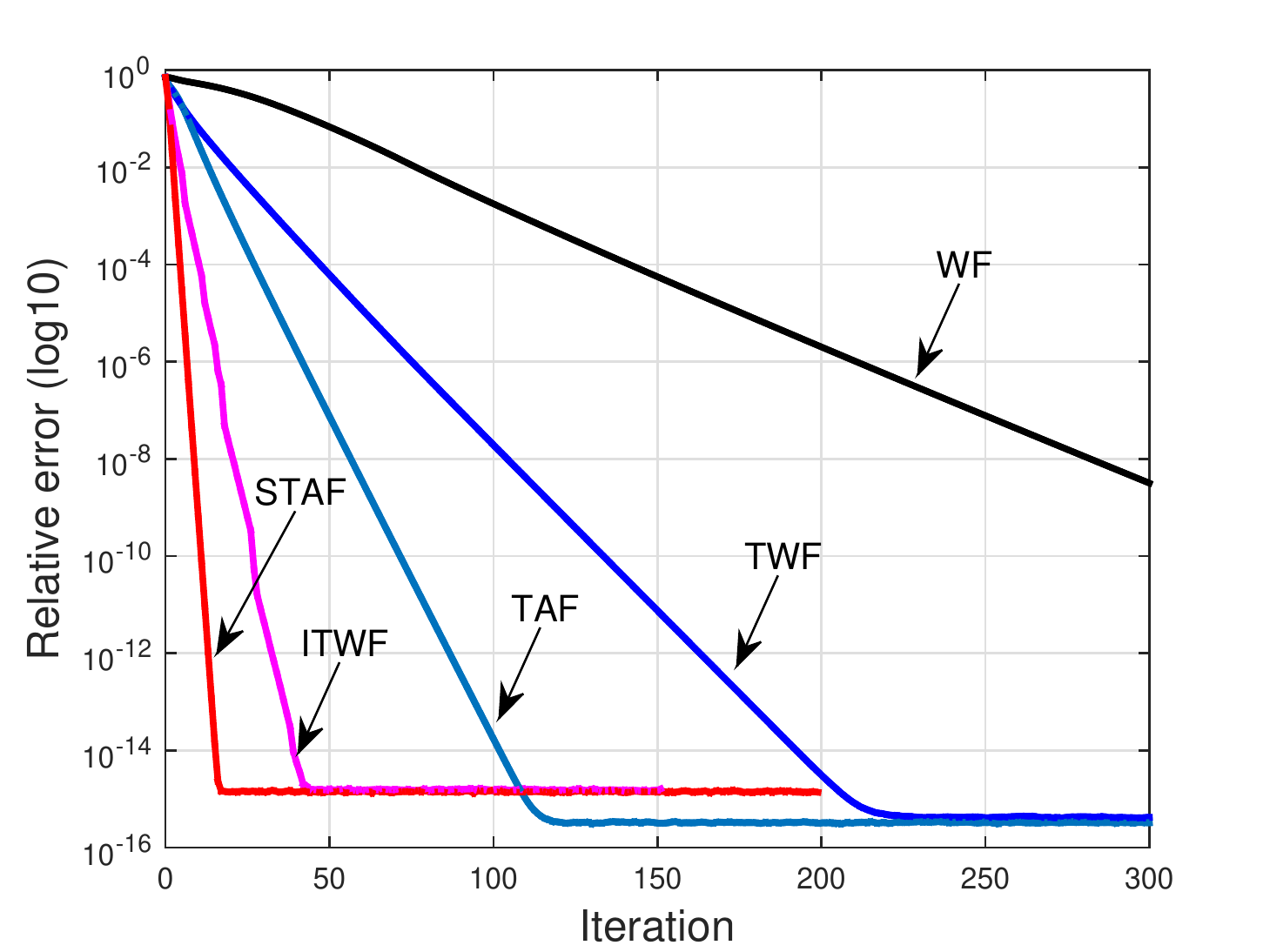}
	\end{subfigure}
	\hspace{-12pt}	
	\begin{subfigure}
	\centering
	\includegraphics[width=.50\textwidth]{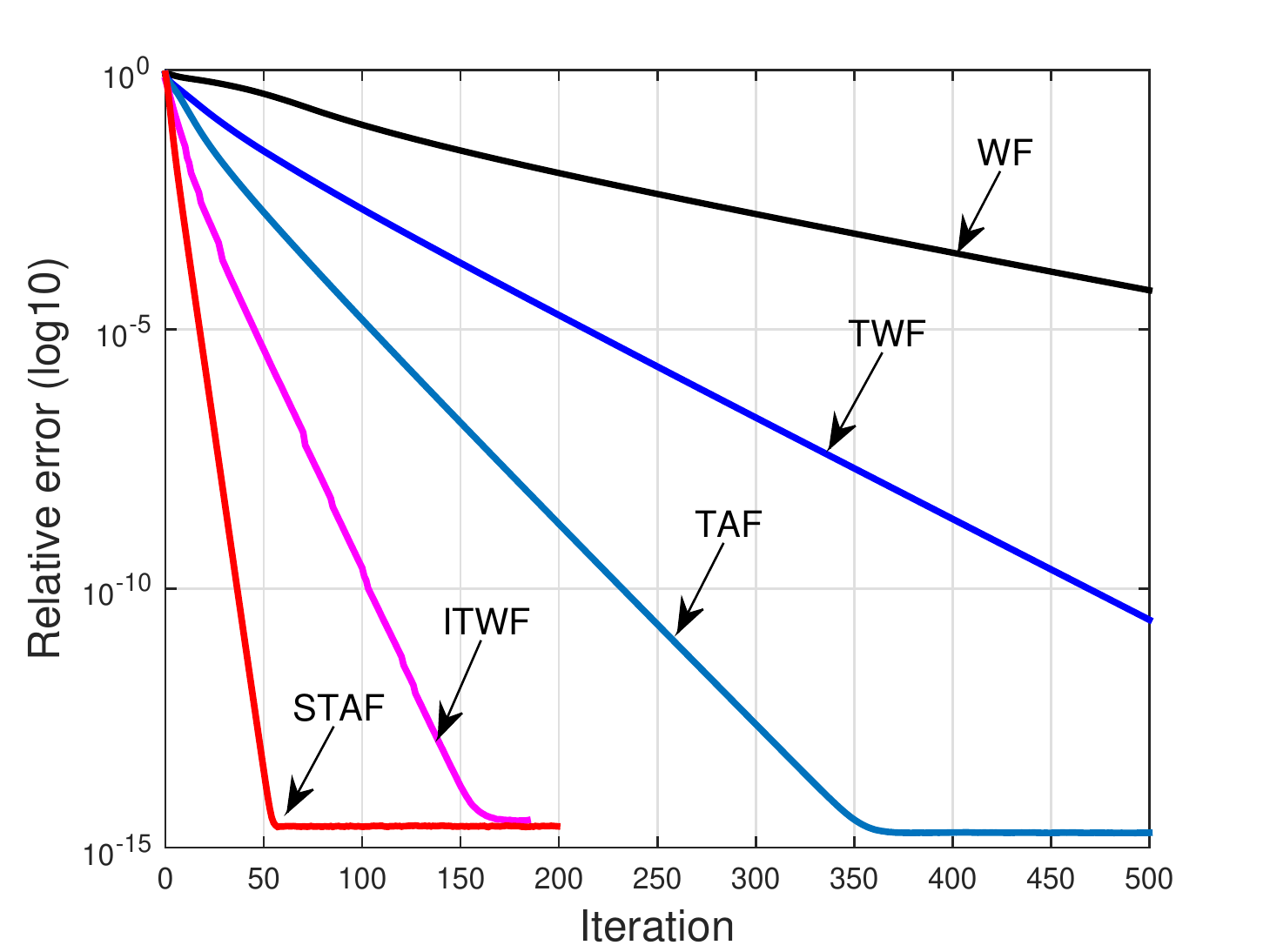} 
	\end{subfigure}
\caption{
Relative error versus iterations using: i) WF~\cite{wf}; ii) TWF~\cite{twf}; iii)~ITAF~\cite{itwf}; iv) TAF~\cite{taf}; and v) STAF under the \emph{same} orthogonality-promoting initialization. Left: Noiseless real-valued Gaussian model with $\bm{x}\sim\mathcal{N}(\bm{0},\bm{I}_n)$, and $\bm{a}_i\sim\mathcal{N}(\bm{0},\bm{I}_n)$; Right: Noiseless complex-valued Gaussian model with $\bm{x}\sim\mathcal{CN}(\bm{0},\bm{I}_n)$, and $\bm{a}_i\sim\mathcal{CN}(\bm{0},\bm{I}_n)$, where $n=1,000$, and $m=5n$. 
}
\label{fig:speed1000}
\vspace{-0pt}
\end{figure}

The second experiment evaluates the  refinement stage of STAF relative to its competing alternatives including those of (T)WF, TAF, and ITWF in a variety of settings. For fairness, all schemes were here initialized using the \emph{same} orthogonality-promoting initialization found using $100$ power iterations, and subsequently applied a number of iterations corresponding to $T=1,000$ data passes. 
First, tests on the noiseless real- and complex-valued Gaussian models were conducted,
with i.i.d. $\bm{a}_i\sim\mathcal{N}(\bm{0},\bm{I}_{1,000})$, $\bm{x}\sim\mathcal{N}(\bm{0},\bm{I}_{1,000})$, and i.i.d. $\bm{a}_i\sim\mathcal{CN}(\bm{0},\bm{I}_{1,000})$, $\bm{x}\sim\mathcal{CN}(\bm{0},\bm{I}_{1,000})$, respectively. 
Figure~\ref{fig:2ndrate} depicts the empirical success rate of all considered schemes with $m/n$ varying by $0.1$ from $1$ to $7$.  
Figure~\ref{fig:speed1000} compares the convergence speed of various schemes in terms of the number of data passes to produce solutions of a given accuracy.
Apparently, starting with the same initialization, STAF outperforms its competing alternatives under both  real-/complex-valued Gaussian models. 
In particular, SGD-based STAF improves in terms of exact recovery and convergence speed
over the state-of-the-art gradient-type TAF, corroborating the benefit of using SGD-type 
solvers to cope with saddle points and local minima of nonconvex optimization~\cite{escaping2015ge,itwf}.  

\begin{figure}[ht!]
	\centering
	\begin{subfigure}
	\centering
	\includegraphics[width=.50\textwidth]{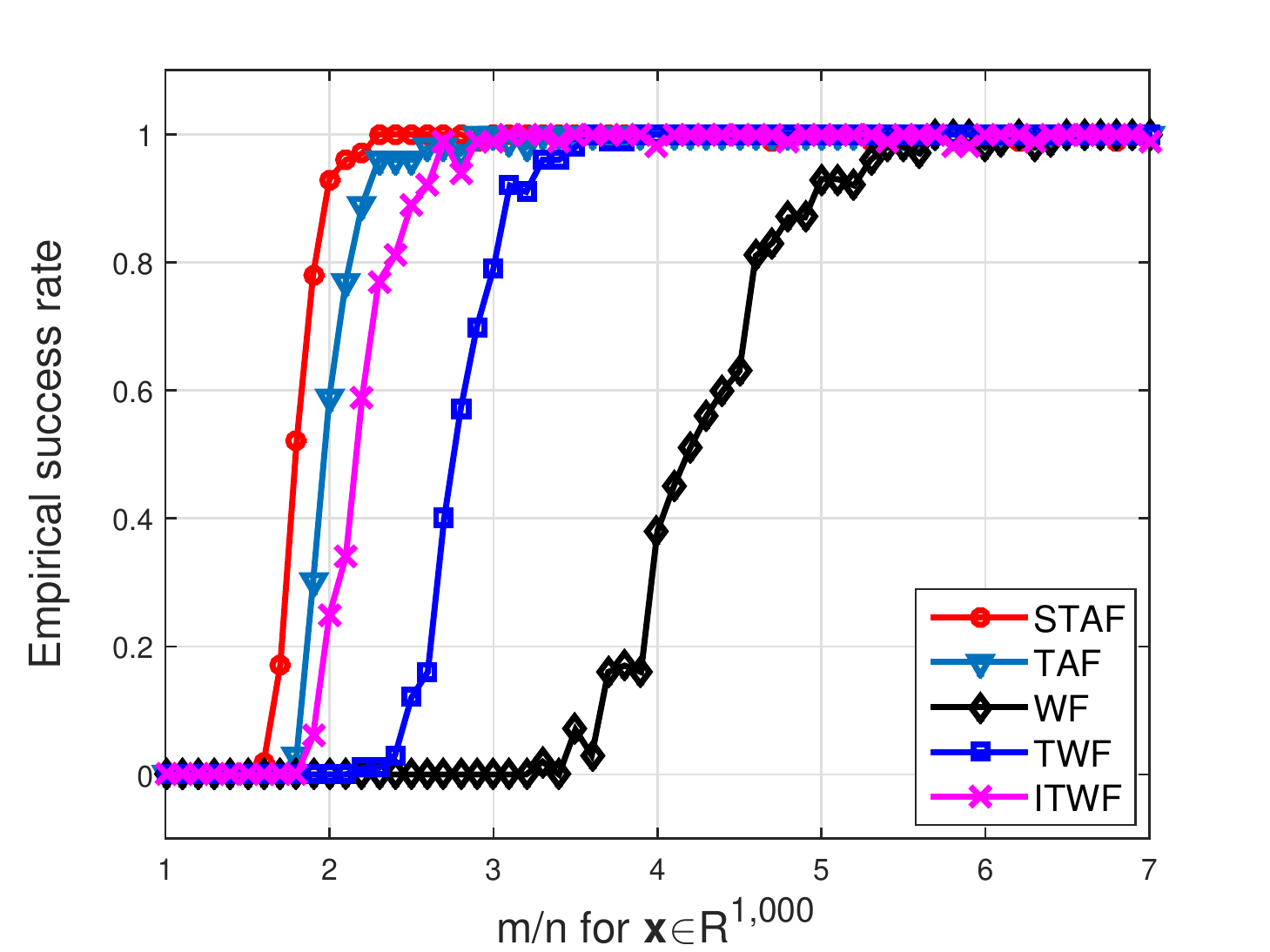}
	\end{subfigure}
	\hspace{-12pt}	
	\begin{subfigure}
	\centering
	\includegraphics[width=.50\textwidth]{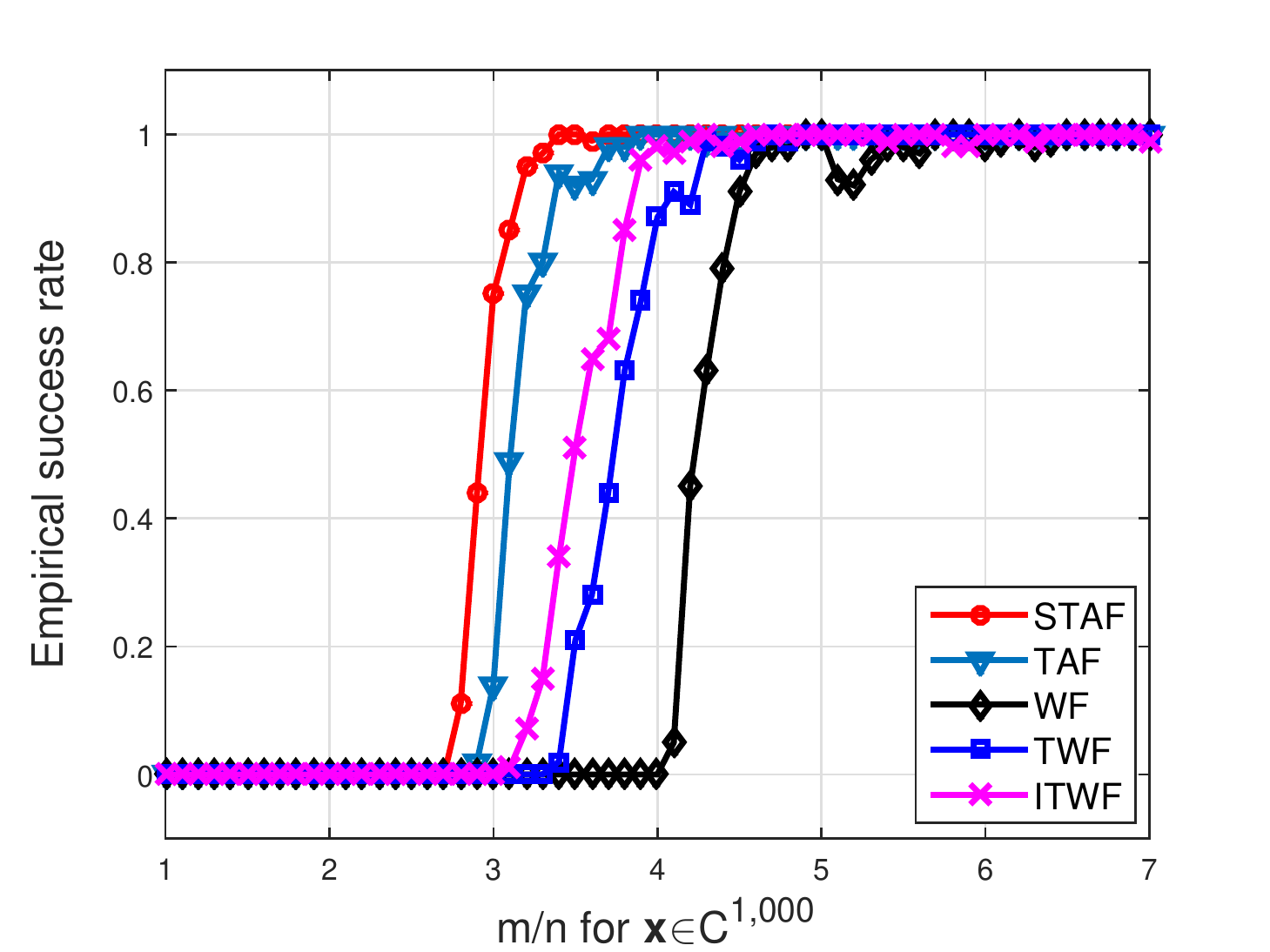} 
	\end{subfigure}
\caption{Empirical success rate for: i) WF~\cite{wf}; ii) TWF~\cite{twf}; iii)~ITAF~\cite{itwf}; iv) TAF~\cite{taf}; and v) STAF with $n=1,000$ and $m/n$ varying $0.1$ from $1$ to $7$. Left: Noiseless real-valued Gaussian model with 
$\bm{x}\sim\mathcal{N}(\bm{0},\bm{I}_n)$, and $\bm{a}_i\sim\mathcal{N}(\bm{0},\bm{I}_n)$; Right: Noiseless complex-valued Gaussian model with 
 $\bm{x}\sim\mathcal{CN}(\bm{0},\bm{I}_n)$, and $\bm{a}_i\sim\mathcal{CN}(\bm{0},\bm{I}_n)$.}
\label{fig:rate}
\vspace{-0pt}
\end{figure}

The previous experiment showed improved performance of STAF under the same initialization. 
Now, we present numerical results comparing different schemes equipped with their own initialization, namely, WF with spectral initialization~\cite{wf}, (I)TWF with truncated spectral initialization~\cite{twf}, as well as TAF with orthogonality-promoting initialization using power iterations~\cite{taf}, and
STAF with VR-OPI. Figure~\ref{fig:rate} demonstrates merits of STAF over its competing alternatives in exact recovery performance on the noiseless real-valued (left) and complex-valued (right) Gaussian model. Specifically in the real case, STAF guarantees exact recovery from about $2.3n$ magnitude-only measurements, which is close to the information-theoretic limit of $m=2n-1$. In addition, STAF achieves about $80\%$ exact recovery rate given the information-limit number of measurements, while all existing except our precursor TAF~\cite{taf} algorithms do not work well in this case. Similar observations hold true when using the complex Gaussian model. 
In comparison, existing alternatives require a few times more measurements to achieve exact recovery. STAF also performs well in the complex case.    
%\textcolor{red}{figure description is missing here......}

\begin{figure}[ht!]
	\centering
	\begin{subfigure}
	\centering
	\includegraphics[width=.50\textwidth]{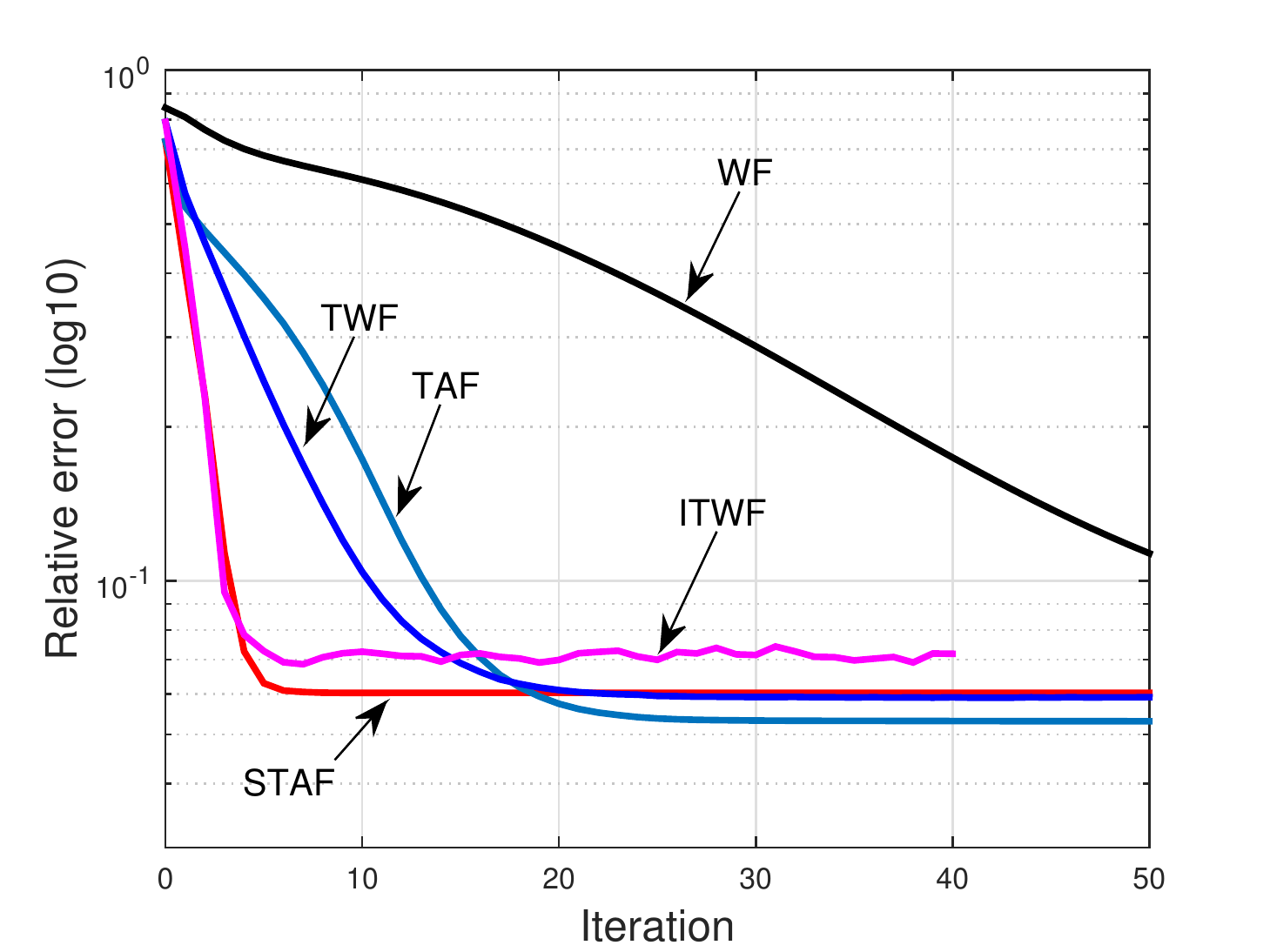}
	\end{subfigure}
	\hspace{-12pt}	
	\begin{subfigure}
	\centering
	\includegraphics[width=.50\textwidth]{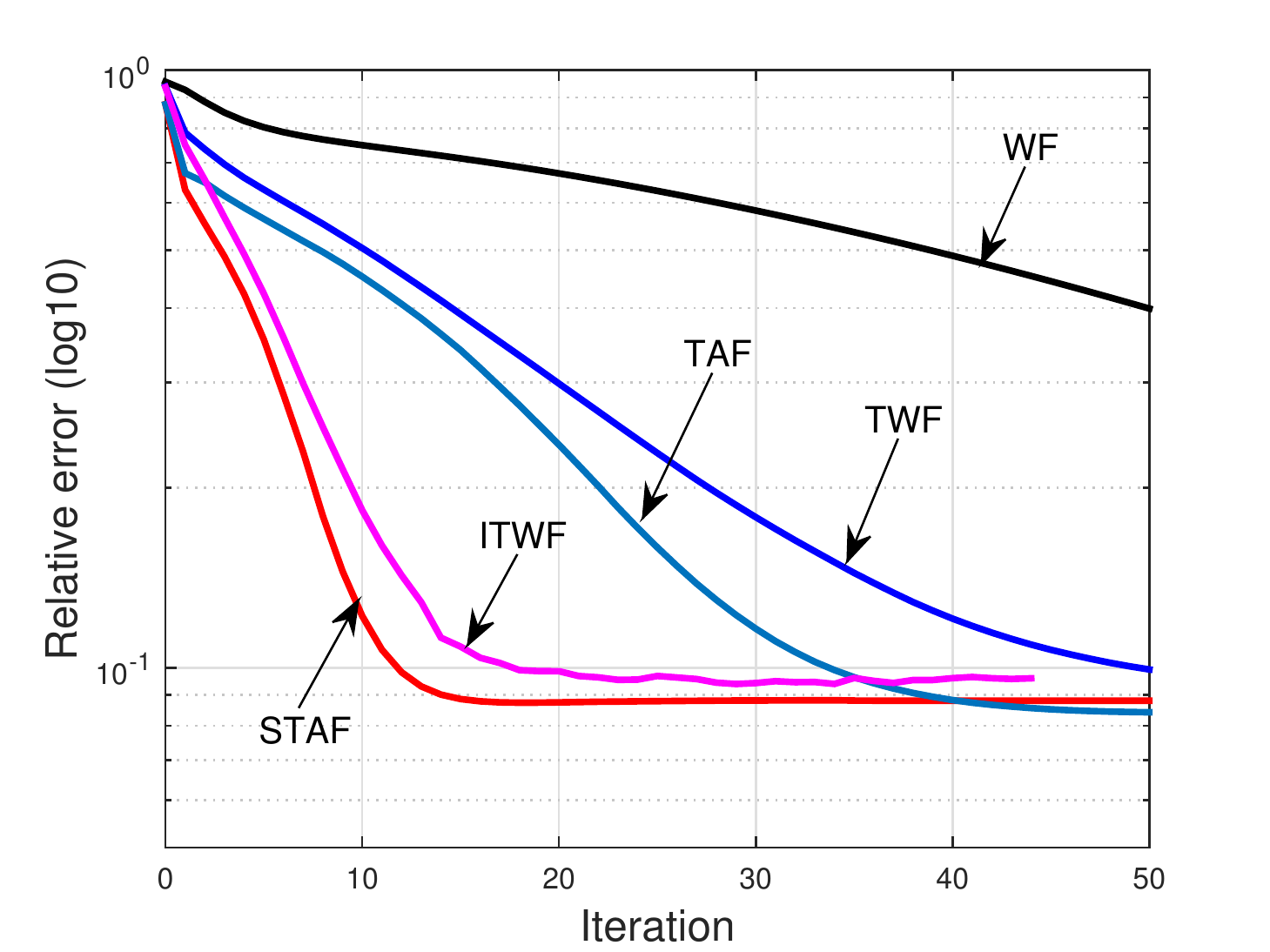} 
	\end{subfigure}
\caption{Relative error versus iterations using: i) WF~\cite{wf}; ii) TWF~\cite{twf}; iii)~ITAF~\cite{itwf}; iv) TAF~\cite{taf}; and v) STAF with $n=1,000$ and $m/n=5$. Left: Noisy real-valued Gaussian model with 
$\bm{x}\sim\mathcal{N}(\bm{0},\bm{I}_n)$, and $\bm{a}_i\sim\mathcal{N}(\bm{0},\bm{I}_n)$; Right: Noisy complex-valued Gaussian model with 
 $\bm{x}\sim\mathcal{CN}(\bm{0},\bm{I}_n)$, and $\bm{a}_i\sim\mathcal{CN}(\bm{0},\bm{I}_n)$.}
\label{fig:noisy}
\vspace{-0pt}
\end{figure}

To demonstrate the robustness of STAF against additive noise, we perform stable phase retrieval under the noisy real-/complex-valued Gaussian model $\psi_i=|\bm{a}_i^\ccalH\bm{x}|+\eta_i$, with $\eta_i\sim\mathcal{N}(\bm{0},\sigma^2\bm{I})$ i.i.d., and $\sigma^2=0.1^2\|\bm{x}\|^2$. Curves in Fig.~\ref{fig:noisy} clearly show near-perfect statistical performance and fast convergence of STAF.

\begin{figure}[ht]
	\centering
	\includegraphics[height=.25\textheight, width=1\textwidth]{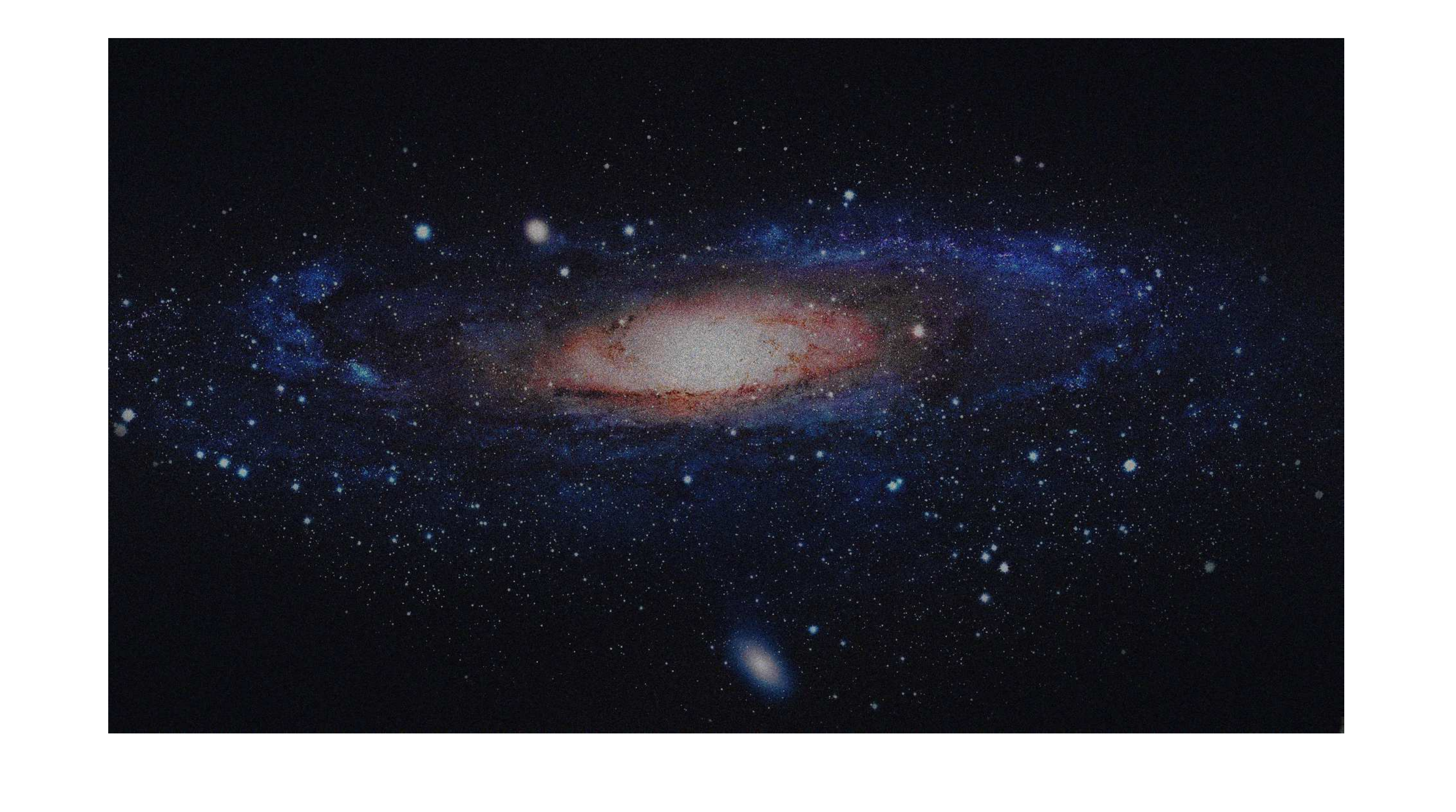}
	\includegraphics[height=.25\textheight, width=1\textwidth]{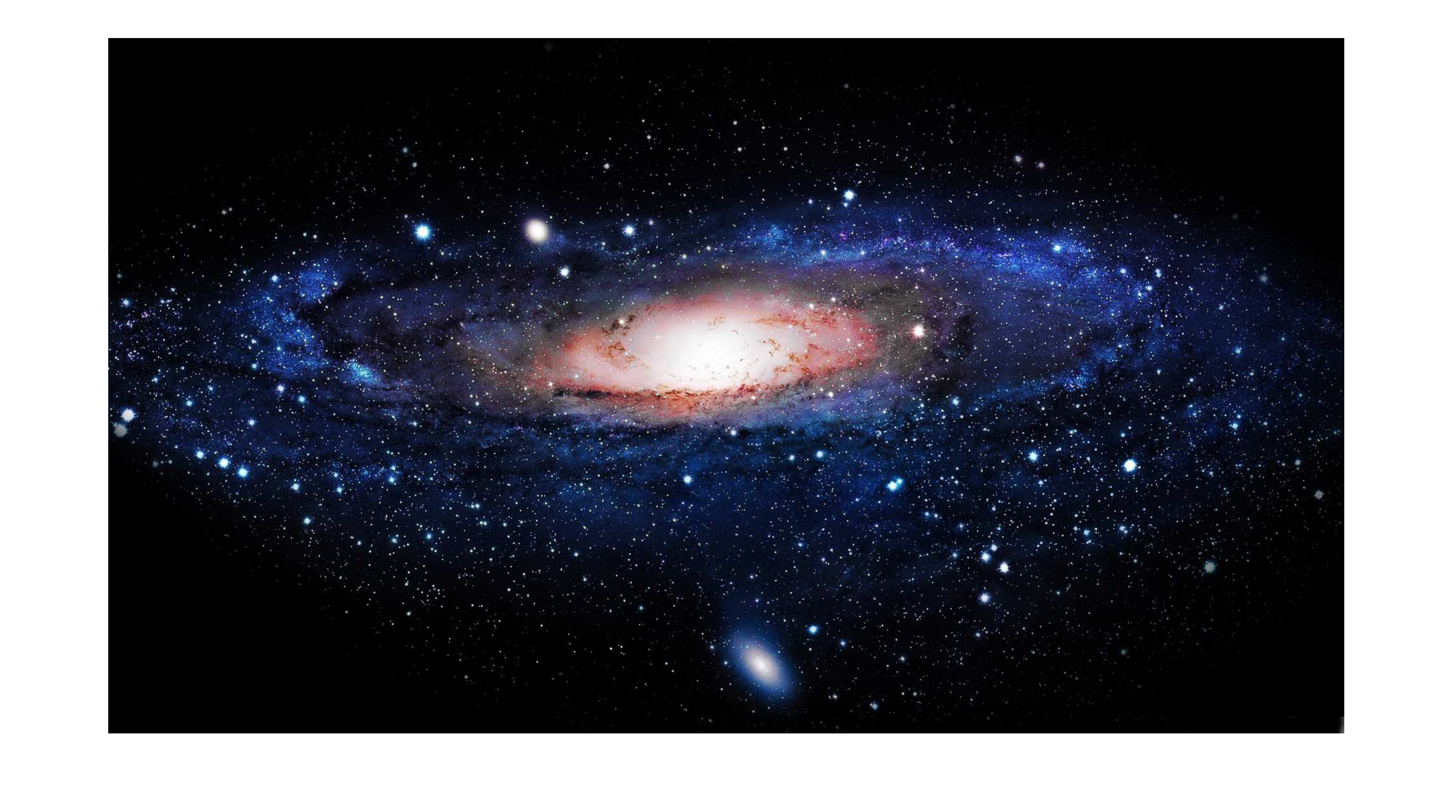}
\caption{
Recovered images after: the variance-reducing orthogonality-promoting initialization stage (top panel), and the STAF refinement stage~(bottom panel)
 on the Milky Way Galaxy image using $K=8$ random masks.
}
\label{fig:milky}
\end{figure}

Finally, to demonstrate the effectiveness and scalability of STAF on real data,
 the Milky Way Galaxy image\footnotemark  \footnotetext{Downloaded from \url{http://pics-about-space.com/milky-way-galaxy.}} 
is considered. The colorful 
image of RGB bands is denoted by $\bm{X}\in\mathbb{R}^{1080\times 1920\times 3}$, 
in which the first two indices encode the pixel location, and the third the color band.
The algorithm was run independently on each of the three RGB images.  
We collected the physically realizable measurements called coded diffraction patterns (CDP) using random masks~\cite{coded}, which have also been used in~\cite{wf,twf,taf,itwf}.
Letting $\bm{x}\in\mathbb{R}^n$ be a vectorization of a certain band of $\bm{X}$,  one has magnitude measurements of the form
\begin{equation}\label{eq:mask}
	\bm{\psi}^{(k)}=\big|\bm{F}\bm{D}^{(k)}\bm{x}\big|,\quad 1\le k\le K
\end{equation}
where $n=1,080\times 1,920=2,073,600$,  
$\bm{F}$ is an $n\times n$ discrete Fourier transform matrix, and $\bm{D}^{(k)}$ is a diagonal matrix whose diagonal entries are sampled uniformly at random from phase delays $\{1,\,-1,\,j,\,-j\}$, with $j$ denoting the imaginary unit. CDP measurements were generated using $K=8$ random masks for a total of $m=nK$  measurements. In this part, since the fast Fourier transform (FFT) can be implemented in $\mathcal{O}(n\log n)$ instead of $\mathcal{O}(n^2)$ operations, the advantage of using STAF with optimal iteration complexity is less pronounced. Hence, instead of processing one quadratic measurement per iteration, a block STAF version processes per iteration $n^2$ measurements associated with one random mask. That is, STAF samples randomly the index $k\in\{1,2,\ldots,K\}$ of masks in~\eqref{eq:mask}, and updates the iterate using all diffraction patterns corresponding to the $k$-th mask. In this case, STAF is able to leverage the efficient implementation of FFT, and  converges fast. 
Figure~\ref{fig:milky} 
displays the recovered images, where the top is obtained after $100$ data passes of VR-OPI iterations, and the bottom is produced by $100$ data passes of STAF iterations refining the initialization. Apparently, the recovered images corroborate the effectiveness of STAF in real-world conditions.

%
%	\begin{figure}[ht]
%	\centering
%	\includegraphics[width=.50\textwidth]{realinit.pdf}
%\caption{
%The error evolution of the  iterates  
%using: i) power method in Algorithm~\ref{alg:pm}; and iii) the developed VR-OPI in Algorithm~\ref{alg:vropi} %for computing the orthogonality-promoting initialization
% on the Milky Way Galaxy image using $K=5$ random masks.
%}
%\label{fig:real}
%	\hspace{-0pt}
%\end{figure}
%

\section{Proof for Theorem~\ref{thm:noiseless}}
\label{pf:noiseless}

Recall from \cite[Thm.1]{taf} that when $m/n$ exceeds some universal constant $c_0>0$, 
the estimate $\bm{z}_0$ returned by the orthogonality-promoting initialization obeys the following with high probability
\begin{equation}
{\rm dist}(\bm{z}_0,\bm{x})\le (1/10) \|\bm{x}\|\label{eq:init}.
\end{equation} 
Along the lines of (T)WF and TAF, 
to prove our Theorem~\ref{thm:noiseless}, it suffices to show that successive STAF iterates $\bm{z}_t$ are on average locally contractive around the planted solution $\bm{x}$, as asserted in the following proposition.  See the Appendix for proof details.
\begin{proposition}[\bf Local error contraction]\label{prop:lec}
Consider the noiseless measurements $\psi_i=|\bm{a}_i^\ccalT\bm{x}|$ with an arbitrary signal $\bm{x}\in\mathbb{R}^n$, and i.i.d. $\bm{a}_i\sim\mathcal{N}(\bm{0},\bm{I}_n)$,~$1\le i \le m$.  
Under the default algorithmic parameters given in Algorithm~\ref{alg:STAF}, there exist universal constants $c_0',\,c_1',\,c_2'>0$, and $\nu>0$, such that with probability at least $1-c_2'm\exp(-c_1'n)$, 
the following holds simultaneously for all $\bm{z}_t$ satisfying \eqref{eq:init}
\begin{equation}\label{eq:lec}
\mathbb{E}_{i_t}\left[{\rm dist}^2(\bm{z}_{t+1},\bm{x})\right]\le \left(1-\frac{\nu}{n}\right){\rm dist}^2(\bm{z}_t,\bm{x})
\end{equation}
provided that $m\ge c_0'n$.

\end{proposition}

Proposition \ref{prop:lec} demonstrates monotonic decrease of the mean-square estimation error:
Once entering a reasonably small-size neighborhood of $\bm{x}$, successive iterates of STAF will be dragged toward $\bm{x}$ at a linear rate. 
Upon establishing the local error contraction property in~\eqref{eq:lec}, taking expectation on both sides of~\eqref{eq:lec} over $i_{t-1}$, and applying Proposition~\ref{prop:lec} again, yields a similar relation for the previous iteration. Continuing this process to reach the initialization $\bm{z}_0$ and appealing to the initialization result in~\eqref{eq:init} collectively, leads to~\eqref{eq:error}, hence completes the proof of Theorem~\ref{thm:noiseless}.

\section{Concluding Remarks}
\label{sec:conclusion}
This paper developed a new linear-time algorithm abbreviated with STAF to solve systems of quadratic equations, and considerably broaden the scope of the state-of-the-art TAF algorithm in~\cite{taf}. Adopting the amplitude-based nonconvex formulation, STAF is a two-stage iterative  algorithm. It
 first adopts an orthogonality-promoting initialization using a stochastic variance reduced gradient algorithm, and subsequently refines the initial estimate via truncated stochastic amplitude-based iterations. STAF was shown capable of recovering any signal from about as many equations as unknowns. In contrast to existing alternatives,  
both stages of STAF achieve optimal iteration and computational complexities that make it attractive to large-scale implementation. Numerical tests involving synthetic data and real images corroborate the merits of STAF in terms of both exact recovery performance and convergence speed over the state-of-the-art algorithms including TAF, (T)WF, and ITWF.  

Pertinent future research directions  
include the possibility of leveraging the orthogonality-promoting initialization in the context of robust phase retrieval and faster semidefinite optimization, and developing suitable gradient regularization rules for other nonconvex optimization tasks. Devising cheap stochastic iterations based solvers for more general nonconvex optimization problems constitutes another meaningful direction.

\section*{Appendix: Proof of Proposition~\ref{prop:lec}}

%\section*{Proof of Proposition~\ref{prop:lec}}

To prove Proposition~\ref{prop:lec}, let us first define the truncated gradient of $\ell(\bm{z})$ as follows
\begin{equation}\label{eq:truncatedgd}
\nabla\ell_{\rm tr}(\bm{z})=\sum_{i=1}^m\left(\bm{a}_i^\ccalT\bm{z}-\psi_i\frac{\bm{a}_i^\ccalT\bm{z}}{|\bm{a}_i^\ccalT\bm{z}|}\right)\bm{a}_i \mathbb{1}_{\big\{\left|\bm{a}_{i}^\ccalT\bm{z}_t\right|\ge \frac{1}{1+\gamma}{\psi_{i}}\big\}}
\end{equation}
which corresponds to the truncated gradient employed by TAF~\cite{taf}. Instrumental in proving the local error contraction in Proposition~\ref{prop:lec}, 
the following lemma adopts a sufficient decrease result from~\cite[Proposition 3]{taf}. The sufficient decrease is a key step in establishing the
local regularity condition \cite{wf,twf,taf}, which suffices to prove linear convergence of iterative optimization algorithms. 
 \begin{proposition}\cite[Proposition 3]{taf}\label{prop:lrc}
 	Consider the noise-free measurements $\psi_i=|\bm{a}_i^\ccalT\bm{x}|$ with i.i.d. $\bm{a}_i\sim\mathcal{N}(\bm{0},\bm{I}_n)$,~$1\le i \le m$, and $\gamma=0.7$. For any fixed  $\epsilon>0$, there exist universal constants $c_0',\,c_1',\,c_2'>0$ such that if $m> c_0'n$,
 	 then the following holds with probability at least $1-c_2'\exp(-c_1'm)$,
\begin{equation}\label{eq:innerproduct}
	\left\langle\bm{h},\frac{1}{m}\nabla	\ell_{\rm tr}(\bm{z})\right\rangle \ge 2\left(1-\zeta_1-\zeta_2-2\epsilon
	\right)\left\|\bm{h}\right\|^2,\quad \bm{h}:=\bm{z}-\bm{x}
\end{equation} 
for all $\bm{x},\,\bm{z}\in\mathbb{R}^n$ such that ${\|\bm{h}\|}/{\|\bm{x}\|}\le {1}/{10}$, where estimates $\zeta_1\approx 0.0782$, and $
\zeta_2\approx 0.2463$.
 \end{proposition}

Now let us turn to the term on the left hand side of \eqref{eq:lec}, which after plugging in the update of $\bm{z}_{t+1}$ in~\eqref{eq:ssgd} or~\eqref{eq:kstaf}, boils down to
\begin{align}
{\rm dist}^2(\bm{z}_{t+1},\bm{x})=&\,\left\|\bm{z}_t-\bm{x}-\mu_t\bigg(\bm{a}_{i_t}^\ccalT\bm{z}_t-\psi_{i_t}\frac{\bm{a}_{i_t}^\ccalT\bm{z}_t}{|\bm{a}_{i_t}^\ccalT\bm{z}_t|}\bigg)\bm{a}_{i_t}\mathbb{1}_{\big\{\left|\bm{a}_{i_t}^\ccalT\bm{z}_t\right|\ge \frac{1}{1+\gamma}{\psi_{i_t}}\big\}}\right\|^2\nonumber\\
=&\,\left\|\bm{h}_t\right\|^2-2\mu_t\bigg(\bm{a}_{i_t}^\ccalT\bm{z}_t-\psi_{i_t}\frac{\bm{a}_{i_t}^\ccalT\bm{z}_t}{|\bm{a}_{i_t}^\ccalT\bm{z}_t|}\bigg)\bm{a}_{i_t}^\ccalT\bm{h}_t \mathbb{1}_{\big\{\left|\bm{a}_{i_t}^\ccalT\bm{z}_t\right|\ge \frac{1}{1+\gamma}{\psi_{i_t}}\big\}}\nonumber\\
&\,+\mu_t^2\bigg(\bm{a}_{i_t}^\ccalT\bm{z}_t-\psi_{i_t}\frac{\bm{a}_{i_t}^\ccalT\bm{z}_t}{|\bm{a}_{i_t}^\ccalT\bm{z}_t|}\bigg)^2\left\|\bm{a}_{i_t}\right\|^2\mathbb{1}_{\big\{\left|\bm{a}_{i_t}^\ccalT\bm{z}_t\right|\ge \frac{1}{1+\gamma}{\psi_{i_t}}\big\}}\label{eq:expanding}
\end{align}
where $\mu_t=\mu>0$ with $i_t\in\{1,\,2,\,\ldots,\,m\}$ sampled uniformly at random~in \eqref{eq:ssgd},   
or $\mu_t=1/\|\bm{a}_{i_t}\|^2$  
with $i_t\in\{1,\,2,\,\ldots,\,m\}$ selected with probability proportional to $\|\bm{a}_{i_t}\|^2$ in~\eqref{eq:kstaf}. 

Consider first the constant step size % with equiprobability
 case in~\eqref{eq:ssgd}. 
 Take the expectation of both sides in \eqref{eq:expanding} with respect to the selection of  index $i_t$ (rather than the data randomness) to obtain
\begin{align}
\mathbb{E}_{i_t}\!\left[{\rm dist}^2(\bm{z}_{t+1},\bm{x})\right]
=&\,\left\|\bm{h}_t\right\|^2-\frac{2\mu}{m}\sum_{i_t=1}^m\bigg(\bm{a}_{i_t}^\ccalT\bm{z}_t-\psi_{i_t}\frac{\bm{a}_{i_t}^\ccalT\bm{z}_t}{|\bm{a}_{i_t}^\ccalT\bm{z}_t|}\bigg)\bm{a}_{i_t}^\ccalT\bm{h}_t \mathbb{1}_{\big\{\left|\bm{a}_{i_t}^\ccalT\bm{z}_t\right|\ge \frac{1}{1+\gamma}{\psi_{i_t}}\big\}}\nonumber\\
&\,+\frac{\mu^2}{m}\sum_{i_t=1}^m\bigg(\bm{a}_{i_t}^\ccalT\bm{z}_t-\psi_{i_t}\frac{\bm{a}_{i_t}^\ccalT\bm{z}_t}{|\bm{a}_{i_t}^\ccalT\bm{z}_t|}\bigg)^2\left\|\bm{a}_{i_t}\right\|^2\mathbb{1}_{\big\{\left|\bm{a}_{i_t}^\ccalT\bm{z}_t\right|\ge \frac{1}{1+\gamma}{\psi_{i_t}}\big\}}\label{eq:expssgd}.
%=&\,\left\|\bm{h}_t\right\|^2-\frac{2\mu}{m}\left\langle \nabla\ell_{\rm tr}(\bm{z}_t),\bm{h}_t\right\rangle +\frac{\mu^2{m}\sum_{i_t=1}^m\bigg(\bm{a}_{i_t}^\ccalT\bm{z}_t-\psi_{i_t}\frac{\bm{a}_{i_t}^\ccalT\bm{z}_t}{|\bm{a}_{i_t}^\ccalT\bm{z}_t|}\bigg)^2\left\\bm{a}_{i_t}\right\|^2\mathbb{1}_{\big\{\left|\bm{a}_{i_t}^\ccalT\bm{z}_t\right|\ge \frac{1}{1+\gamma}{\psi_{i_t}}\big\}}\nonumber.
\end{align}
Now the task reduces to upper bounding the terms on the right hand side of \eqref{eq:expssgd}. Note from \eqref{eq:truncatedgd} that by means of $\nabla\ell_{\rm tr}(\bm{z}_t)$, the second term in~\eqref{eq:expssgd} can be re-expressed as follows
\begin{align}\label{eq:second}
-\frac{2\mu}{m}\sum_{i_t=1}^m\bigg(\bm{a}_{i_t}^\ccalT\bm{z}_t-\psi_{i_t}\frac{\bm{a}_{i_t}^\ccalT\bm{z}_t}{|\bm{a}_{i_t}^\ccalT\bm{z}_t|}\bigg)\bm{a}_{i_t}^\ccalT\bm{h}_t \mathbb{1}_{\big\{\left|\bm{a}_{i_t}^\ccalT\bm{z}_t\right|\ge \frac{1}{1+\gamma}{\psi_{i_t}}\big\}}&=-\frac{2\mu}{m}\left\langle \nabla\ell_{\rm tr}(\bm{z}_t),\bm{h}_t\right\rangle\nonumber\\
&\le- {4\mu}\left(1-\zeta_1-\zeta_2-2\epsilon
	\right)\left\|\bm{h}\right\|^2
\end{align}
where the inequality follows from Proposition~\ref{prop:lrc}. 
Regarding the last term in~\eqref{eq:expssgd}, since for the i.i.d. real-valued Gaussian $\bm{a}_i$'s, $\max_{i_t\in[m]}\|\bm{a}_{i_t}\|\le 2.3n$ holds with probability at least $1-m{\rm e}^{-n/2}$~\cite{taf}, and also $\mathbb{1}_{\big\{\left|\bm{a}_{i_t}^\ccalT\bm{z}_t\right|\ge \frac{1}{1+\gamma}{\psi_{i_t}}\big\}}\le 1$, then the next holds with high probability
\begin{align}
\frac{\mu^2}{m}\sum_{i_t=1}^m\bigg(\bm{a}_{i_t}^\ccalT\bm{z}_t-\psi_{i_t}\frac{\bm{a}_{i_t}^\ccalT\bm{z}_t}{|\bm{a}_{i_t}^\ccalT\bm{z}_t|}\bigg)^2\left\|\bm{a}_{i_t}\right\|^2\mathbb{1}_{\big\{\left|\bm{a}_{i_t}^\ccalT\bm{z}_t\right|\ge \frac{1}{1+\gamma}{\psi_{i_t}}\big\}}&\le \frac{2.3n\mu^2}{m}\sum_{i_t=1}^m\left(\left|\bm{a}_{i_t}^\ccalT\bm{z}_t\right|-\left|\bm{a}_{i_t}^\ccalT\bm{x}\right|\right)^2\nonumber\\
%\mathbb{1}_{\big\{\left|\bm{a}_{i_t}^\ccalT\bm{z}_t\right|\ge \frac{1}{1+\gamma}{\psi_{i_t}}\big\}}
&\le \frac{2.3n\mu^2}{m}\sum_{i_t=1}^m\left(\bm{a}_{i_t}^\ccalT\bm{z}_t-\bm{a}_{i_t}^\ccalT\bm{x}\right)^2\nonumber\\
&\le  \frac{2.3n\mu^2}{m}\bm{h}_t^\ccalT\bm{A}^\ccalT\bm{A}\bm{h}_t\nonumber\\
&\le 2.3(1+\delta)\mu^2n\left\|\bm{h}_t\right\|^2\label{eq:third}
\end{align}
in which the second inequality comes from $(|\bm{a}_{i_t}^\ccalT\bm{z}_t|-|\bm{a}_{i_t}^\ccalT\bm{x}|)^2\le (\bm{a}_{i_t}^\ccalT\bm{z}_t-\bm{a}_{i_t}^\ccalT\bm{x})^2$, and the
last inequality arises due to the fact that $\lambda_{\max}(\bm{A}^\ccalT\bm{A})\le (1+\delta)m$ holds with probability at least $1-c_2'\exp(-c_1'n\delta^2)$, provided that $m\ge c_0'n\delta^{-2}$ for some universal constant $c_0',\,c_1',\,c_2'>0$~\cite[Theorem 5.39]{chap2010vershynin}.  

Substituting \eqref{eq:second} and \eqref{eq:third} into \eqref{eq:expssgd} establishes that
\begin{equation}
\label{eq:finalssgd}
\mathbb{E}_{i_t}\!\left[{\rm dist}^2(\bm{z}_{t+1},\bm{x})\right]\le \left[1- 4\mu\left(1-\zeta_1-\zeta_2-2\epsilon
	\right)+2.3(1+\delta)\mu^2n
\right]\left\|\bm{h}_t\right\|^2
\end{equation}
holds with probability exceeding $1-c_2m\exp(-c_1n)$ provided that $m\ge c_0n$, where $c_0\ge c_0'\delta^{-2}$. To obtain legitimate estimates for the step size, fixing $\epsilon,\,\delta>0$ to be sufficiently small constants, say e.g., $0.01$, then using \eqref{eq:finalssgd}, $\mu$ can be chosen such that $4(0.98-\zeta_1-\zeta_2
	)-2.42\mu n> 0$, yielding
\begin{equation}
0<\mu<\frac{4(0.98-\zeta_1-\zeta_2
	)}{2.42n}\approx \frac{1.0835}{n}:=\frac{\mu_0}{n}.
\end{equation}
Plugging $\mu=c_3/n$ for some $0<c_3\le \mu_0$ into \eqref{eq:finalssgd}, gives rise to
\begin{equation}
\label{eq:sfinal}
\mathbb{E}_{i_t}\!\left[{\rm dist}^2(\bm{z}_{t+1},\bm{x})\right]\le \left(1- \frac{\nu}{n}
\right){\rm dist}^2(\bm{z}_{t},\bm{x})
\end{equation}
for $\nu:= 4c_3(1-\zeta_1-\zeta_2-2\epsilon)-2.3c_3^2(1+\delta)\le \nu_0:=0.1139$, where the equality holds at the maximum step size $\mu=\mu_0$, 
 hence concluding the proof of Proposition~\ref{prop:lec} for the constant step size case.

Now let us turn to the case of a time-varying step size. Specifically, let $\mu_t=1/\|\bm{a}_{i_t}\|^2$, and $i_t$ be sampled at random from the set $\{1,\,2,\,\ldots,\,m\}$ with probability $\|\bm{a}_{i_t}\|^2/\sum_{i_t=1}^m\|\bm{a}_{i_t}\|^2=\|\bm{a}_{i_t}\|^2/\|\bm{A}\|_{F}^2$~\cite{strohmer2009}.  
Taking the expectation of both sides in \eqref{eq:expanding} over $i_t$ gives rise to
\begin{align}
\mathbb{E}_{i_t}\!\left[{\rm dist}^2(\bm{z}_{t+1},\bm{x})\right]
=&\,\left\|\bm{h}_t\right\|^2-2\sum_{i_t=1}^m\frac{\|\bm{a}_{i_t}\|^2}{\|\bm{A}\|_{F}^2}\frac{1}{\|\bm{a}_{i_t}\|^2}\bigg(\bm{a}_{i_t}^\ccalT\bm{z}_t-\psi_{i_t}\frac{\bm{a}_{i_t}^\ccalT\bm{z}_t}{|\bm{a}_{i_t}^\ccalT\bm{z}_t|}\bigg)\bm{a}_{i_t}^\ccalT\bm{h}_t \mathbb{1}_{\big\{\left|\bm{a}_{i_t}^\ccalT\bm{z}_t\right|\ge \frac{1}{1+\gamma}{\psi_{i_t}}\big\}}\nonumber\\
&\,+\sum_{i_t=1}^m\frac{\|\bm{a}_{i_t}\|^2}{\|\bm{A}\|_{F}^2}\frac{1}{\|\bm{a}_{i_t}\|^2}\bigg(\bm{a}_{i_t}^\ccalT\bm{z}_t-\psi_{i_t}\frac{\bm{a}_{i_t}^\ccalT\bm{z}_t}{|\bm{a}_{i_t}^\ccalT\bm{z}_t|}\bigg)^2\mathbb{1}_{\big\{\left|\bm{a}_{i_t}^\ccalT\bm{z}_t\right|\ge \frac{1}{1+\gamma}{\psi_{i_t}}\big\}}\label{eq:kfirst}.
%=&\,\left\|\bm{h}_t\right\|^2-\frac{2\mu}{m}\left\langle \nabla\ell_{\rm tr}(\bm{z}_t),\bm{h}_t\right\rangle +\frac{\mu^2{m}\sum_{i_t=1}^m\bigg(\bm{a}_{i_t}^\ccalT\bm{z}_t-\psi_{i_t}\frac{\bm{a}_{i_t}^\ccalT\bm{z}_t}{|\bm{a}_{i_t}^\ccalT\bm{z}_t|}\bigg)^2\left\\bm{a}_{i_t}\right\|^2\mathbb{1}_{\big\{\left|\bm{a}_{i_t}^\ccalT\bm{z}_t\right|\ge \frac{1}{1+\gamma}{\psi_{i_t}}\big\}}\label{eq:kfirst}.
\end{align}
Consider random $\bm{A}:=[\bm{a}_1~\cdots~\bm{a}_m]^\ccalT$ with
i.i.d. rows $\bm{a}_i\sim\mathcal{N}(\bm{0},\bm{I}_n)$, and any fixed $\sigma>0$. Then, by means of Bernstein-type inequality~\cite[Proposition 5.16]{chap2010vershynin}, $\big|\frac{1}{mn}\|\bm{A}\|_F^2-1\big|=\big|\frac{1}{mn}\sum_{i,j}a_{i,j}^2-1\big|\le \sigma$ holds with probability at least $1-2\exp(-mn\sigma^2/8)$. 
Therefore, %due to $\big(\bm{a}_{i_t}^\ccalT\bm{z}_t-\psi_{i_t}\frac{\bm{a}_{i_t}^\ccalT\bm{z}_t}{|\bm{a}_{i_t}^\ccalT\bm{z}_t|}\big)\bm{a}_{i_t}^\ccalT\bm{h}_t\mathbb{1}_{\big\{\left|\bm{a}_{i_t}^\ccalT\bm{z}_t\right|\ge \frac{1}{1+\gamma}{\psi_{i_t}}\big\}}\ge 0$, $1\le  i_t\le m$,
 the second term on the right hand side of \eqref{eq:kfirst} can be bounded as follows
\begin{align}
	&-\frac{2}{\|\bm{A}\|_{F}^2}\sum_{i_t=1}^m\bigg(\bm{a}_{i_t}^\ccalT\bm{z}_t-\psi_{i_t}\frac{\bm{a}_{i_t}^\ccalT\bm{z}_t}{|\bm{a}_{i_t}^\ccalT\bm{z}_t|}\bigg)\bm{a}_{i_t}^\ccalT\bm{h}_t \mathbb{1}_{\big\{\left|\bm{a}_{i_t}^\ccalT\bm{z}_t\right|\ge \frac{1}{1+\gamma}{\psi_{i_t}}\big\}}\nonumber\\
	&\le-\frac{2}{(1+\sigma)mn} \sum_{i_t=1}^m\bigg(\bm{a}_{i_t}^\ccalT\bm{z}_t-\psi_{i_t}\frac{\bm{a}_{i_t}^\ccalT\bm{z}_t}{|\bm{a}_{i_t}^\ccalT\bm{z}_t|}\bigg)\bm{a}_{i_t}^\ccalT\bm{h}_t \mathbb{1}_{\big\{\left|\bm{a}_{i_t}^\ccalT\bm{z}_t\right|\ge \frac{1}{1+\gamma}{\psi_{i_t}}\big\}}\nonumber\\
	&\le -\frac{4m}{(1+\sigma)mn}\left(1-\zeta_1-\zeta_2-2\epsilon
	\right)\left\|\bm{h}\right\|^2\nonumber\\
	&\le -\frac{4}{(1+\sigma)n}\left(1-\zeta_1-\zeta_2-2\epsilon
	\right)\left\|\bm{h}\right\|^2\label{eq:ksecond}
\end{align}
where the second inequality follows from Proposition \ref{prop:lrc}, and the last inequality from the fact that $m\ge c_0 n$.
Concerning the last term on the right hand side of~\eqref{eq:kfirst}, one obtains that
%it can be easily verified that 
%\begin{equation}
%	\sum_{i_t=1}^m\left|\left\langle\bm{a}_{i_t},\bm{h}\right\rangle\right|^2\ge \frac{\|\bm{h}\|^2}{\|\bm{A}^{-1}\|^2}
%\end{equation}
%holds for all $\bm{h}\in\mathbb{R}^n$. Then, using the fact that $\|\bm{A}\|_F^2=\sum_{i_t=1}^m\|\bm{a}_{i_t}\|^2$, one can rewrite the third term as
%the following holds with probability at least $1-c_2'\exp(-c_1'n\delta^2)$ 
\begin{align}
	&\sum_{i_t=1}^m\frac{\|\bm{a}_{i_t}\|^2}{\|\bm{A}\|_{F}^2}\frac{1}{\|\bm{a}_{i_t}\|^2}\bigg(\bm{a}_{i_t}^\ccalT\bm{z}_t-\psi_{i_t}\frac{\bm{a}_{i_t}^\ccalT\bm{z}_t}{|\bm{a}_{i_t}^\ccalT\bm{z}_t|}\bigg)^2\mathbb{1}_{\big\{\left|\bm{a}_{i_t}^\ccalT\bm{z}_t\right|\ge \frac{1}{1+\gamma}{\psi_{i_t}}\big\}}\nonumber\\
	&=\frac{1}{\|\bm{A}\|_{F}^2}\sum_{i_t=1}^m\left(\left|\bm{a}_{i_t}^\ccalT\bm{z}_t\right|-\left|\bm{a}_{i_t}^\ccalT\bm{x}\right|\right)^2\mathbb{1}_{\big\{\left|\bm{a}_{i_t}^\ccalT\bm{z}_t\right|\ge \frac{1}{1+\gamma}{\psi_{i_t}}\big\}}\nonumber\\
	&\le \frac{1}{\|\bm{A}\|_{F}^2}\sum_{i_t=1}^m\left(\bm{a}_{i_t}^\ccalT\bm{z}_t-\bm{a}_{i_t}^\ccalT\bm{x}\right)^2\nonumber\\
	&\le \frac{1}{\|\bm{A}\|_{F}^2}\bm{h}_{t}^\ccalT\bm{A}^\ccalT\bm{A}\bm{h}_t\nonumber\\
&\le \frac{(1+\delta)m}{(1-\sigma)mn}\left\|\bm{h}_t\right\|^2\nonumber\\
	&\le \frac{(1+\delta)}{(1-\sigma)n}\left\|\bm{h}_t\right\|^2\label{eq:kthird}
\end{align}
which holds with high probability as soon as
$m\ge c_0n\ge c_0'\delta^{-2}n$. 

Putting results in \eqref{eq:kfirst}, \eqref{eq:ksecond}, and \eqref{eq:kthird} together, one establishes that the following holds
\begin{equation}
\mathbb{E}_{i_t}\!\left[{\rm dist}^2(\bm{z}_{t+1},\bm{x})\right]\le \left[1-\frac{4}{(1+\sigma)n}\left(1-\zeta_1-\zeta_2-2\epsilon
	\right)+\frac{(1+\delta)}{(1-\sigma)n}
\right]\left\|\bm{h}_t\right\|^2\label{eq:kfinal}
\end{equation}
 with probability at least $1-c_2m\exp(-c_1n)$ provided that $m\ge c_0n$. Hence, one can set in this case $$\nu:=\frac{4}{(1+\sigma)n}\left(1-\zeta_1-\zeta_2-2\epsilon
	\right)-\frac{(1+\delta)}{(1-\sigma)n}.$$ 
 Taking without loss of generality $\delta,\,\sigma,\,\epsilon$ to be $0.01$, and substituting the estimates of $\zeta_1,\,\zeta_2$ into \eqref{eq:kfinal}, one arrives at $\nu=1.5758$ to deduce that
\begin{equation}
\label{eq:kstep}
\mathbb{E}_{i_t}\!\left[{\rm dist}^2(\bm{z}_{t+1},\bm{x})\right]\le\left(1-\frac{1.5758}{n}\right){\rm dist}^2(\bm{z}_t,\bm{x})
\end{equation}
which holds with high probability as soon as $m\ge c_0n$, establishing the local error contraction property of the truncated Kaczmarz iterations in~\eqref{eq:kstaf}, as claimed in Proposition \ref{prop:lec}. 

Combining the results in \eqref{eq:sfinal} and \eqref{eq:kstep}, we proved the local error contraction property in Proposition~\ref{prop:lec} of the two STAF variants
under both constant and time-varying step sizes. 

\bibliographystyle{IEEEtran}

\bibliography{apower}
%\bibliography{/Users/wang/Dropbox/apower}

%\bibliography{/Users/dtc-sa-wang/Dropbox/apower}

%\balance
\end{document}